\documentclass[prb,twocolumn,superscriptaddress,amsmath,amssymb,floatfix]{revtex4-2}

\usepackage[T1]{fontenc}
\usepackage[utf8]{inputenc}
\usepackage{lmodern}
\usepackage[english]{babel}
\input{glyphtounicode}
\AtBeginDocument{\lefthyphenmin=2\righthyphenmin=3}
\usepackage{graphicx}
\graphicspath{{}}
\usepackage{dcolumn}
\usepackage{bm}
\usepackage{xcolor}
\usepackage[colorlinks=true,linkcolor=blue,citecolor=blue,urlcolor=blue]{hyperref}

\newcommand{\bk}{\mathbf k}
\newcommand{\hex}{h_{\rm ex}}
\newcommand{\heff}{h_{\rm eff}}
\newcommand{\DEc}{\Delta E_{\rm cut}}
\newcommand{\Fc}{F_{\rm cut}}
\newcommand{\Fm}{F_{\rm cut}^{\max}}
\newcommand{\Zmaj}{Z_{\rm maj}}
\newcommand{\Zmin}{Z_{\rm min}}
\newcommand{\Smaj}{\Sigma_{\rm maj}}
\newcommand{\Smin}{\Sigma_{\rm min}}
\newcommand{\Zedge}{Z_{\rm edge}}
\newcommand{\Lf}{\mathcal{L}}
\newcommand{\Rf}{\mathcal{R}}

\begin{document}

\title{Altermagnetic spin splitting across the Slater--Mott crossover}

\author{Philipp Werner}
\email{philipp.werner@unifr.ch}
\affiliation{Department of Physics, University of Fribourg, 1700 Fribourg, Switzerland}

\author{Francesco Petocchi}
\affiliation{Department of Physics, University of Fribourg, 1700 Fribourg, Switzerland}

\author{Sayed Ali Akbar Ghorashi}
\email{ghorashi@sas.upenn.edu}
\affiliation{Department of Chemistry, University of Pennsylvania, Philadelphia, Pennsylvania 19104, USA}

\date{\today}

\begin{abstract}
Momentum-dependent spin splitting distinguishes altermagnets from conventional collinear antiferromagnets, yet many candidate materials host strong electronic correlations. Clarifying how this splitting evolves as interactions reshape the electronic structure is essential to connect the underlying magnetic order with spectroscopic measurements. Employing Hartree--Fock theory and dynamical mean-field theory (DMFT), we systematically trace the spin-resolved spectral functions of a $g$-wave altermagnetic Hubbard model across the Slater--Mott crossover and under finite electron doping. At half filling, the DMFT spin splitting reaches a maximum at intermediate coupling and shrinks at stronger interactions, even though the staggered magnetic moment approaches saturation. This nonmonotonic behavior reflects the competition between increasing spin--sublattice locking and dynamical band renormalization. At strong coupling, electron doping rapidly screens the low-frequency exchange field, and the splitting decreases approximately in proportion to the ordered moment, in sharp contrast to the Hartree--Fock prediction, where it remains close to saturation. The single-particle spectra display a pronounced redistribution of spectral weight between the two spin branches and into a ladder of spin-polaron sidebands, while at strong coupling the first moments of the removal spectra closely track the Hartree--Fock dispersions. We show that the unequal weights of the two spin branches originate from the frequency dependence of the local self-energy near the gap edge. Our findings demonstrate that the magnitude of the ordered altermagnetic moment alone does not govern the energy splittings and spectral intensities of the quasiparticles.
\end{abstract}

\maketitle

\section{Introduction}
Altermagnets host momentum-dependent spin splitting while preserving zero net magnetization. This spin splitting originates from the interplay of crystalline symmetries and collinear magnetic order, and persists even without spin--orbit coupling~\cite{Smejkal2022a,Smejkal2022b,hayami2019momentum}. While photoemission experiments in MnTe and CrSb confirmed this characteristic band dispersion~\cite{Krempasky2024,Reimers2024,ARPES2,ARPES5}, electronic correlations obscure the relation between the magnetic order and the single-particle excitations. Specifically, a nearly saturated local moment can coexist with strong dynamical bandwidth reduction, spectral-weight suppression, and a frequency-dependent exchange field. Hubbard models with sublattice-dependent hopping amplitudes~\cite{Das2024,Giuli2025} provide a minimal platform to explore this correlated regime.

Correlation effects are especially pronounced in altermagnetic Mott insulators~\cite{Guo2023,Wei2025}, where photoemission injects a hole that propagates through an ordered spin background. In usual antiferromagnets, this motion generates spin polarons with narrow bandwidths, suppressed coherent weights, and incoherent higher-energy excitations~\cite{Brinkman1970,Kane1989,Martinez1991}. Recent work has extended this framework to altermagnets. Daghofer \textit{et al.} revealed pronounced spin-selective spectral-weight transfer, where the coherence of individual branches depends sensitively on the magnetic background and hopping parameters~\cite{Daghofer2026}. Lanzini \textit{et al.} established that spin-polaron bandwidths and splittings undergo substantial suppression relative to static Hartree--Fock predictions~\cite{Lanzini2026}. In parallel, combined Hartree--Fock and dynamical mean-field theory (DMFT)~\cite{Georges1996} calculations uncovered interaction- and doping-driven band reconstruction in altermagnetic Hubbard systems~\cite{DelRe2025}, while spin-dependent quasiparticle lifetimes and correlation-induced modifications of the splitting have been examined in both model and material settings~\cite{Leraand2026,Park2025}.

Previous model investigations focused predominantly on $d$-wave altermagnets~\cite{Das2024,Giuli2025,Daghofer2026,Lanzini2026,DelRe2025,Leraand2026}, whereas prototypical materials such as MnTe and CrSb~\cite{jungwirth2025altermagnetism,jungwirth2026symmetry} realize $g$-wave symmetry. In this work, we track the spin splitting of a $g$-wave Hubbard model across the Slater--Mott crossover, spanning weak to strong coupling at half filling and under finite electron doping. We find that the low-energy spin splitting evolves nonmonotonically with interaction strength, peaking at intermediate coupling despite a staggered magnetization that grows monotonically at low temperature. Within Hartree--Fock, the splitting saturates once the exchange field quenches the intersublattice hopping, so that the Bloch states of each spin reside predominantly on one sublattice (spin--sublattice locking). In contrast, DMFT reveals that local quantum fluctuations introduce a dynamical band renormalization that eventually outweighs the increasing locking and suppresses the splitting in the Mott regime. Remarkably, despite this low-energy suppression, the first moments of the DMFT removal spectra track the Hartree--Fock dispersion at strong coupling. Electron doping reveals another distinct role of strong correlations. The low-frequency exchange field decays far more rapidly than its Hartree value, rendering the altermagnetic splitting significantly more fragile than static mean-field theory suggests. Finally, we show that the two dominant spin branches near the gap edge carry markedly asymmetric spectral weights despite being governed, in the strongly locked limit, by the same local majority self-energy.

The paper is structured as follows. In Sec.~\ref{sec:model}, we introduce the $g$-wave Hubbard model and define the diagnostic metrics that disentangle spin--sublattice locking from dynamical band renormalization. Sections~\ref{sec:half} and \ref{sec:doped} present the interaction dependence at half filling and the doping evolution, focusing on the DMFT results. Section~\ref{sec:spectra} analyzes the spin-resolved spectral distributions, branch weights, and spectral moments, while Sec.~\ref{sec:discussion} provides a brief summary. 
Hartree--Fock benchmarks, weak-coupling Slater spectra, the effective string-chain mapping, and numerical implementation details are presented in the appendices.

\section{Model and methods} \label{sec:model}
Following the two-sublattice geometry of Refs.~\cite{Roig2024,Gondolf2025}, we consider a square Bravais lattice hosting two sites per unit cell, labeled $A$ at $(0,0)$ and $B$ at $(\tfrac12,\tfrac12)$, described by the Hubbard Hamiltonian
\begin{equation}
H=\sum_{\bk\sigma}c^\dagger_{\bk\sigma}h(\bk)c_{\bk\sigma} +U\sum_i n_{i\uparrow}n_{i\downarrow}.
\label{eq:H}
\end{equation}
Here, the fermionic spinor is $c_{\bk\sigma}=(c_{A\bk\sigma},c_{B\bk\sigma})^T$, and the noninteracting single-particle Hamiltonian reads $h(\bk)=\varepsilon_0(\bk)\tau_0+\varepsilon_x(\bk)\tau_x+\varepsilon_z(\bk)\tau_z$, where the Pauli matrices $\tau_i$ act in sublattice space and the dispersion components are
\begin{equation}
\begin{aligned}
\varepsilon_0(\bk)&=t_1(\cos k_x+\cos k_y),\\
\varepsilon_x(\bk)&=t_x\cos\tfrac{k_x}{2}\cos\tfrac{k_y}{2},\\
\varepsilon_z(\bk)&=t_z\sin k_x\sin k_y(\cos k_x-\cos k_y).
\end{aligned}
\label{eq:eps}
\end{equation}
The intersublattice hopping $\varepsilon_x$ hybridizes sublattices $A$ and $B$, while $\varepsilon_z$ imparts opposite dispersions to the two sublattices. The function $\varepsilon_z$ is invariant under $C_4$ and changes sign under the axial and diagonal mirrors, giving the $A_{2g}$ (or $g$-wave) form factor. Throughout this work, we fix $t_1=0$, $t_x=3.4$, and $t_z=1.5$, yielding a bare noninteracting bandwidth $W_0=6.8$. The average electron density per site is denoted by $n$, with half filling corresponding to $n=1$.

We focus on the collinear altermagnetic ground state characterized by $n_A=n_B=n$ and a staggered magnetization $m_s \equiv m_A=-m_B$, where $m_\alpha=n_{\alpha\uparrow}-n_{\alpha\downarrow}$. At the Hartree--Fock level, the effective Hamiltonian for spin projection $\sigma=\pm1$ is
\begin{equation}
H_\sigma^{\rm HF}=\frac{Un}{2}\tau_0+\varepsilon_x\tau_x +(\varepsilon_z-\sigma\hex)\tau_z,
\label{eq:HHF}
\end{equation}
with the mean-field exchange field given by
\begin{equation}
\hex=\frac{Um_s}{2}.
\label{eq:hex}
\end{equation}
Diagonalizing Eq.~\eqref{eq:HHF} yields the Hartree--Fock energy eigenvalues $E_{\sigma\nu}^{\rm HF}=Un/2+\nu R_\sigma$, where $\nu=\pm1$ labels the band branch and $R_\sigma=[\varepsilon_x^2+(\varepsilon_z-\sigma\hex)^2]^{1/2}$.

\begin{figure}[t]
\includegraphics[width=\columnwidth]{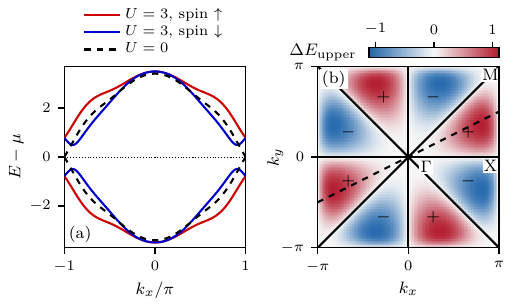}
\caption{Hartree--Fock bands and nodes at half filling. (a) Bands along the cut $k_y=k_x/2$ for $U=0$ (dashed) and $U=3$ (solid, $m_s=0.524$). The dotted line marks the chemical potential. The ordered-state gap along the cut is smaller than the zone-boundary separation $2\hex=1.57$. (b) Signed splitting of the upper band at $U=3$. The axes, diagonals, and zone boundary are spin degenerate. The dashed line marks the cut used throughout the paper.}
\label{fig:bands}
\end{figure}

At half filling (chemical potential $\mu=U/2$), the band structure displays several notable characteristics. Both hopping terms $\varepsilon_x$ and $\varepsilon_z$ vanish identically along the Brillouin-zone boundary, producing a noninteracting sublattice degeneracy [Fig.~\ref{fig:bands}(a)]. Setting $t_1=0$ pins this degenerate manifold to $E=\mu$ and enforces particle--hole symmetry. In the altermagnetically ordered Hartree--Fock state, the zone-boundary bands split to $E-\mu=\pm\hex$, opening an insulating gap across the Fermi level. Concurrently, the magnetic order lifts the spin degeneracy throughout the interior of the zone. The signed spin splitting $\Delta E_\nu=\nu(R_\uparrow-R_\downarrow)$ vanishes wherever $\varepsilon_z\hex=0$, pinning the $g$-wave nodal lines along $k_x=0$, $k_y=0$, and $k_x=\pm k_y$, with additional nodal degeneracies along the zone boundary [Fig.~\ref{fig:bands}(b)]. The sign of the splitting alternates between neighboring octants, and inverting the staggered moment $m_s\to-m_s$ reverses the sign globally. The momentum trajectory $k_y=k_x/2$ [dashed line in Fig.~\ref{fig:bands}(b)] intersects nodal lines solely at the $\Gamma$ point and the zone boundary. Elsewhere, it traverses two octants of identical sign with $\varepsilon_z\le0$, ensuring that the splitting along this path maintains a uniform sign. In the upper band, the spin-up branch lies above the spin-down branch, while in the lower band this ordering is inverted.

We benchmark the DMFT simulations against this altermagnetic Hartree--Fock solution at temperature $T=0.01$. The mean-field competition with a uniform ferromagnet and the stability against spiral states are discussed in Appendix~\ref{app:hf}.

Within single-site DMFT~\cite{Metzner1989,MullerHartmann1989,Georges1996}, the electronic self-energy $\Sigma$ is spatially local, but frequency dependent, and diagonal in sublattice space. The magnetic point-group symmetry dictates $\Sigma_{A\uparrow}=\Sigma_{B\downarrow}\equiv \Smaj$ and $\Sigma_{A\downarrow}=\Sigma_{B\uparrow}\equiv \Smin$. At half filling, particle--hole conjugation combined with a single-site translation yields the relation $\Smaj(\omega)=U-\Smin(-\omega)^*$ (with energy $\omega$ measured relative to $\mu=U/2$), which links the local renormalization factors at opposing band edges. Here, spatial translation interchanges sublattices $A$ and $B$, canceling the sign inversion of $\varepsilon_z$ under particle--hole transformation. 

We solve the effective impurity problem using hybridization expansion continuous-time quantum Monte Carlo (CT-QMC)~\cite{Werner2006} at $T=0.01$, and a zero-temperature exact-diagonalization (ED)~\cite{Caffarel1994} solver. A configuration-interaction truncation~\cite{Zgid2011,Zgid2012} extends the ED solution to $N_s=161$ impurity and bath sites. The CT-QMC solver is used for the weak-coupling regime and doping scans, while the ED solver allows to resolve the sharp gap-edge excitations and spin-polaron sidebands at half filling. The analytic continuation of the CT-QMC data and the ED bath discretization scheme are detailed in Appendix~\ref{app:ed}. We refer to the occupied part of the spectral function, which is measured by photoemission, as the removal spectrum. At $T=0$, it is nonzero only for $\omega\le0$ (relative to $\mu$) and given by $\sum_m|\langle m|c_{\bk\sigma}|\Psi_0\rangle|^2\delta(\omega+E_m-E_0)$, where $|\Psi_0\rangle$ is the interacting ground state and the $|m\rangle$ denote states with one electron removed. The unoccupied part of the spectrum at $\omega>0$ is the addition spectrum. While all ED calculations are performed at $T=0$, the low-frequency exchange field defined below is evaluated on the imaginary frequency axis.

To quantify the magnitude of the altermagnetic splitting, we evaluate its maximum along the cut $k_y=k_x/2$ [dashed line in Fig.~\ref{fig:bands}(b)], denoted $\DEc$, tracking the band nearest the gap or the Fermi level. Within Hartree--Fock, this splitting depends on interaction strength and carrier filling solely through the exchange field, $\DEc^{\rm HF}=\Fc(\hex)$, where
\begin{equation}
\Fc(h)=\max_{\text{cut}} \left|\sqrt{\varepsilon_x^2+(\varepsilon_z-h)^2} -\sqrt{\varepsilon_x^2+(\varepsilon_z+h)^2}\right|.
\label{eq:F}
\end{equation}
In the weak-field limit, $\Fc(h)$ grows linearly with $h$. In contrast, when $h\gg|\varepsilon_x|,|\varepsilon_z|$, the intersublattice hopping $\varepsilon_x$ is suppressed, and the Bloch eigenstates of the lower band become strongly polarized onto sublattice $A$ for spin up and sublattice $B$ for spin down (and vice versa in the upper band). We call this type of polarization spin--sublattice locking. In the fully spin--sublattice locked regime, each spin projection disperses exclusively via the same-sublattice hopping $\pm\varepsilon_z$, so that the splitting at each momentum saturates at $2|\varepsilon_z(\bk)|$. Hence, the Hartree--Fock splitting saturates at $\Fm=\max_{\rm cut}2|\varepsilon_z|=2.26$ (within $2\%$ of the global Brillouin-zone maximum), with the half-saturation field $h_{1/2}\simeq0.85$ (Appendix~\ref{app:hf}).

To systematically compare the static mean-field reference with DMFT, we define the effective low-frequency exchange field
\begin{equation}
\heff=|\operatorname{Re}\delta\Sigma(\omega=0)|, \qquad \delta\Sigma=\frac{\Smaj-\Smin}{2},
\label{eq:heff}
\end{equation}
along with two dimensionless diagnostic ratios:
\begin{equation}
\Lf=\frac{\Fc(\heff)}{\Fm},\qquad \Rf=\frac{\DEc}{\Fc(\heff)}.
\label{eq:LR}
\end{equation}
The locking factor $\Lf$ measures how close the static reference is to the saturated splitting, while the renormalization factor $\Rf$ quantifies the remaining dynamical reduction, yielding the exact decomposition $\DEc=\Rf\Lf\Fm$. Disentangling $\Lf$ and $\Rf$ helps to clarify the underlying physics. 

In CT-QMC calculations, we determine $\heff$ at the lowest Matsubara frequency $\omega_0=\pi T$ ($T=0.01$). For a direct comparison, we evaluate the $T=0$ ED self-energy, which is known analytically from its Lehmann representation, at the same imaginary frequency $i\omega_0$. All other ED results, including the moments, gaps, splittings, spectra, and local renormalization factors, are obtained on the real-frequency axis at $T=0$. At half filling, $\omega=0$ lies within the insulating gap where $\operatorname{Im}\Sigma$ vanishes, so that $\operatorname{Re}\delta\Sigma(\omega)$ remains smooth even in the Mott state (the paramagnetic self-energy pole at $\omega=0$ is shifted by the magnetic exchange into the Hubbard bands). Consequently, Matsubara and real-frequency values of $\heff$ coincide to better than $10^{-3}$. The physical gap-edge states reside at finite energy, where $|\text{Re}\,\delta\Sigma|$ is slightly reduced (by $4$--$7\%$ for $U=6$--$15$). Thus, at half filling, $\heff$ provides an effective reference field rather than the exact exchange field acting on the band-edge states.

Finally, we characterize the frequency dependence of the self-energy via the local renormalization factors
\begin{equation}
Z_\alpha(\omega)= \left[1-\partial_\omega\operatorname{Re}\Sigma_\alpha(\omega)\right]^{-1}, \qquad \alpha={\rm maj},{\rm min},
\label{eq:Zalpha}
\end{equation}
which are well-defined in regions of weak damping $\operatorname{Im}\Sigma_\alpha\simeq0$. The spectral weight of any lattice pole is determined by these local factors weighted by the wave-function sublattice composition (Sec.~\ref{sec:spectra}). 

\begin{figure*}[ht]
\includegraphics[width=\textwidth]{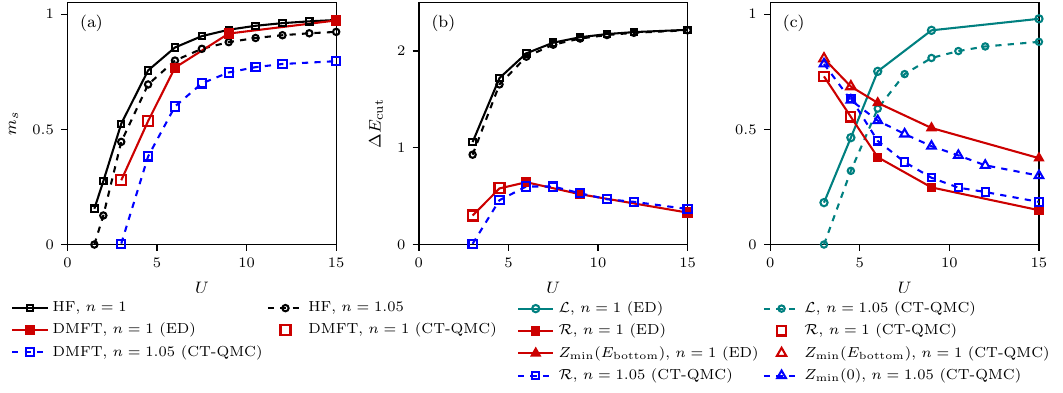}
\caption{Interaction dependence at half filling and at $n=1.05$. (a) Staggered moment and (b) maximum spin splitting along $k_y=k_x/2$. Black: Hartree--Fock, red: half-filled DMFT, blue: doped DMFT results. Solid and dashed lines distinguish $n=1$ and $n=1.05$, respectively. (c) Splitting renormalization $\Rf$ (squares), locking factor $\Lf$ (green), and the local renormalization factor [Eq.~\eqref{eq:Zalpha}] at the band edge (triangles). At half filling, the latter is $\Zmin(E_{\rm bottom})$ at the bottom of the upper band, while at $n=1.05$, it is $\Zmin(0)$. At $U=3$, the $n=1.05$ state is paramagnetic and $\Rf$ is undefined. HF and CT-QMC use $T=0.01$, while ED uses $T=0$. The impurity solver is indicated in the labels.}
\label{fig:dmftU}
\end{figure*}

\section{Interaction-driven evolution of the spin splitting} \label{sec:half}
Figure~\ref{fig:dmftU} contrasts the interaction-dependent staggered magnetization with the resulting spin splitting. In Hartree--Fock theory, both $m_s$ and $\DEc$ increase monotonically with $U$. In DMFT, by contrast, the spin splitting exhibits a pronounced nonmonotonic profile, peaking near $U\approx W_0$ both at half filling and in the electron-doped regime. At half filling, as $U$ increases from $6$ to $15$, the staggered moment rises from $m_s=0.768$ to $0.973$, while the maximal spin splitting drops from $\DEc=0.65$ to $0.33$. Although the magnetic moment is virtually saturated at strong coupling, the low-energy spin splitting is suppressed by nearly a factor of two. Hence, within a correlated dynamical treatment, a larger altermagnetic order parameter does not imply a larger observable spin splitting.

This nonmonotonic evolution arises from the competing trends of the two factors in Eq.~\eqref{eq:LR}. In the weak-coupling regime, the exchange field is smaller than the intersublattice hopping $|\varepsilon_x|$, leaving the Bloch states strongly mixed between sublattices. At $U=3$, the locking factor is only $\Lf=0.18$, and increasing $U$ amplifies the splitting by enhancing the exchange field. At strong coupling, however, spin--sublattice locking is essentially complete, reaching $\Lf=0.93$ at $U=9$ and $\Lf=0.98$ at $U=15$. In this regime, further increases in the exchange field produce negligible gains in $\Fc(\heff)$, resulting in a saturation of the Hartree--Fock splitting. In DMFT, however, the dynamical renormalization factor $\Rf$ decreases monotonically [Fig.~\ref{fig:dmftU}(c)], falling from $0.70$ at $U=3$ down to $0.15$ at $U=15$. Once spin--sublattice locking saturates, this dynamical reduction dominates the interaction dependence. The maximum splitting emerges near $U=6\approx W_0$, where locking is largely established while the dynamical renormalization remains moderate. Across the investigated interaction window, the half-filled ED values of $\Rf$ follow an approximate $1/U$ scaling ($U\Rf\simeq2.1$--$2.3$, see also Appendix~\ref{app:strings}). 

The nonmonotonic behavior of $\DEc$ is accompanied by a qualitative reshaping of the excitation spectrum. At $U=3$ and $4.5$, the states flanking the insulating gap are dispersive Slater-like bands characterized by reduced gaps and modest weight renormalization. Here, CT-QMC and ED agree almost quantitatively, for staggered moments within $2\%$ and band splittings within $7\%$. These small differences mainly reflect the finite ED bath and uncertainties related to the analytic continuation of the CT-QMC data. The temperature difference ($T=0.01$ versus $T=0$) plays a minor role because $T$ is far below the gap. At large interactions, each Hubbard band reorganizes into a narrow edge band with $g$-wave spin splitting, accompanied by an extensive ladder of spin-polaron sidebands (Sec.~\ref{sec:spectra}). For $U=15$, the DMFT single-particle gap is $9.50$, substantially smaller than the Hartree--Fock value of $12.64$, even though both theories yield nearly identical first moments for the spectral distributions. The weak-coupling Slater spectra are detailed in Appendix~\ref{app:half}, and the strong-coupling Mott features are examined in Sec.~\ref{sec:spectra}.

\begin{figure*}[t]
\includegraphics[width=\textwidth]{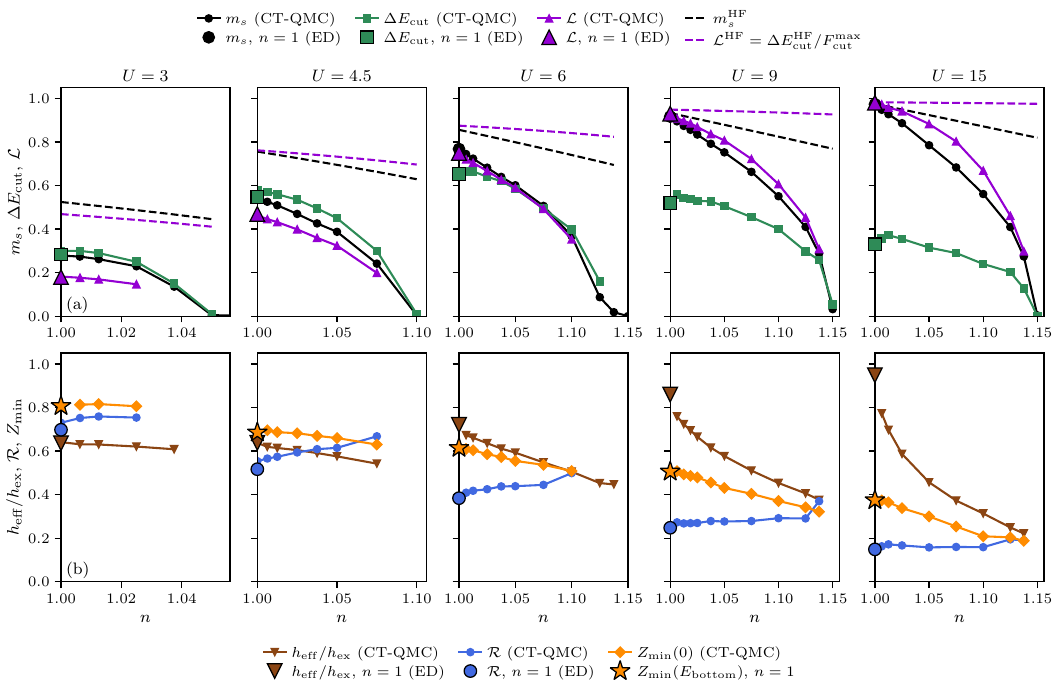}
\caption{Doping dependence across the Slater--Mott crossover for $U=3$, $4.5$, $6$, $9$, and $15$. (a) Ordered moment $m_s$ (black), maximal splitting $\DEc$, and locking factor $\Lf$ (purple). Dashed lines: Hartree--Fock moment $m_s^{\rm HF}$ and locking factor $\Lf^{\rm HF}=\DEc^{\rm HF}/\Fm$ at the same $U$ and $n$. (b) Dynamical reduction factors, which all equal one in Hartree--Fock: the fraction $\heff/\hex$ of the exchange field that survives at low frequency (brown), the renormalization $\Rf=\DEc/\Fc(\heff)$ (blue), and the local renormalization factor at the band edge (orange), given in the doped metal by $\Zmin(0)$ and at half filling by $\Zmin(E_{\rm bottom})$ (stars, CT-QMC for $U\le4.5$ and ED for $U\ge6$). Lines with small symbols: CT-QMC ($T=0.01$). Large triangles, circles, and squares at $n=1$: ED ($T=0$, $N_s=33$ for $U=3$, $61$ otherwise). In each column, $\DEc=\Rf\Lf\Fm$ with $\Fm=2.26$. $\Lf$, $\Rf$, and $\Zmin$ are shown only for $m_s\ge0.15$, where $\Rf$ can be measured.}
\label{fig:zall}
\end{figure*}

This nonmonotonic splitting persists upon carrier doping [blue curves in Fig.~\ref{fig:dmftU}]. At $n=1.05$, the splitting peaks at approximately $0.60$ near $U=6$--$7.5$ before decaying at stronger coupling, whereas the staggered moment continues to grow monotonically. Increasing the Hubbard interaction thus separates the ground-state magnetic order from the low-energy quasiparticle splitting in both the correlated insulator and the doped metallic state.

\section{Doping and dynamical exchange-field screening} \label{sec:doped}
In the strong-coupling regime, Hartree--Fock predicts that the altermagnetic spin splitting is largely immune to doping-induced moment reduction [dashed lines in Fig.~\ref{fig:zall}(a); see also Fig.~\ref{fig:collapse}(b) for the full doping range]. At $U=15$, increasing electron filling from $n=1$ to $1.15$ decreases the Hartree--Fock moment $m_s^{\rm HF}$ from $0.975$ to $0.819$, yet the splitting $\DEc^{\rm HF}$ shifts only marginally from $2.22$ to $2.20$. Because the mean-field exchange field remains far above the half-saturation threshold $h_{1/2}$ throughout this doping range, the splitting stays pinned to its locking ceiling. In DMFT, by contrast, the splitting remains approximately constant only at very weak doping. Beyond $n\approx1.025$, the splitting decays in tandem with the staggered moment, and long-range order disappears near $n=1.15$ for both $U=9$ and $U=15$.

\begin{figure*}[t]
\includegraphics[width=\textwidth]{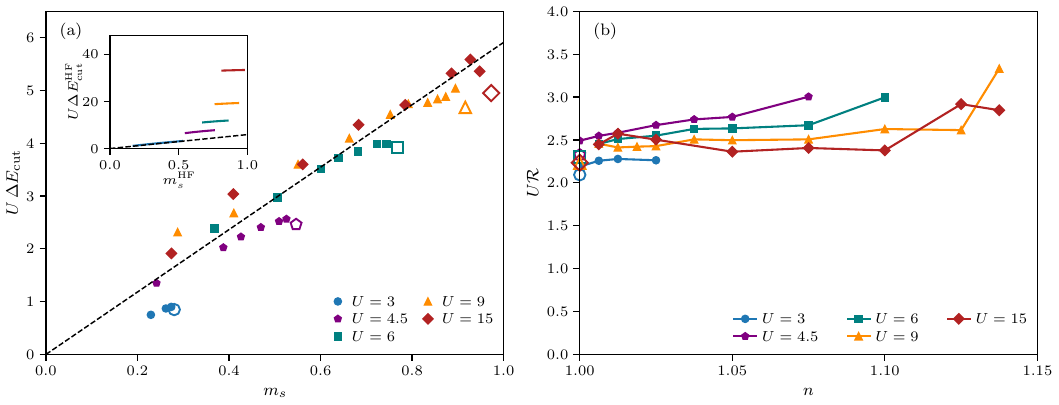}
\caption{(a) Spin splitting versus ordered moment along the CT-QMC doping scans ($T=0.01$, filled symbols, $m_s\ge0.15$), together with the half-filled ED values (open symbols). Dashed line: $U\DEc=5.9\,m_s$, fitted to the doped data for $U=6$, $9$, and $15$. Inset: Hartree--Fock at the same interactions for $n=1$--$1.15$, where the splitting remains close to its saturated value as the moment decreases. (b) Renormalization of the splitting multiplied by $U$ along the doping scans (lines, CT-QMC) and at half filling (open symbols, ED).}
\label{fig:mscollapse}
\end{figure*}

Figure~\ref{fig:zall} analyzes this doping response using the factorization $\DEc=\Rf\Lf\Fm$. The top row presents quantities directly comparable to Hartree--Fock, while the bottom row shows dynamical reduction factors, which equal one in Hartree--Fock. The staggered moment decreases approximately linearly at low doping, and the extrapolation of this CT-QMC result to half filling matches the zero-temperature ED value within $0.01$ for all interactions. The locking factor $\Lf$ decreases with doping at all $U$, and in the Mott regime ($U\gtrsim 6$) roughly in proportion to the moment, because the low-frequency exchange field decays much faster than the bare Hartree value $\hex=Um_s/2$ [brown curves in Fig.~\ref{fig:zall}(b)]. As long as $\heff$ remains substantially above $h_{1/2}$, spin--sublattice locking remains nearly saturated, explaining the initial plateau of the band splitting at very small doping in the Mott regime. Meanwhile, the dynamical renormalization factor $\Rf$ exhibits a weak doping dependence, showing a moderate increase only near the magnetic phase boundary \footnote{The modest upturn of $\Rf$ at high doping reflects the fact that $\heff$ is evaluated at $\omega=0$ rather than at the finite energy of the maximally split states.}. Consequently, at strong coupling, the spin splitting decays more gradually than $\Lf$, approximately tracking the decline of the staggered magnetization.

We call the suppression of $\heff$ relative to $\hex$ the dynamical screening of the exchange field. At weak coupling ($U=3$ and $4.5$), the screening ratio $\heff/\hex$ decreases only slightly with doping. At strong coupling ($U=9$ and $15$), however, $\heff/\hex$ drops between half filling and $n=1.00625$ and continues a steep decline upon further doping [Fig.~\ref{fig:zall}(b)]. Thus, carrier-induced screening of the altermagnetic exchange field, encoded in the frequency dependence of the self-energy, is immediately activated upon doping of the Mott insulator.

Figure~\ref{fig:mscollapse}(a) illustrates the relationship between the spin splitting and the staggered moment across all doping trajectories. Along each scan, the product $U\DEc/m_s$ remains approximately constant, so that the splitting and the magnetization decrease roughly in parallel. In the Mott regime, the data approximately collapse onto $U\DEc\simeq5.9\,m_s$, with slight upward deviations near the end of the ordered region, where $\Rf$ rises. Hartree--Fock exhibits no such relation (inset). The approximate data collapse is the result of two separate trends. First, the scaled renormalization $U\Rf$ is nearly constant across doping and interaction [Fig.~\ref{fig:mscollapse}(b)], yielding $\Rf\approx2.5/U$ from the Slater to the Mott regime. Second, the decrease of the locking factor $\Lf$ approximately tracks that of the staggered moment.

At weak coupling, DMFT and Hartree--Fock behave qualitatively similarly [Figs.~\ref{fig:zall} and \ref{fig:collapse}(b)]. For $U=3$ and $4.5$, dynamic screening remains weak, and spin--sublattice locking is unsaturated even at half filling. In both theories, the splitting responds approximately linearly to the exchange field, decaying alongside the moment from the onset of doping.

The stability of $\Rf$ in Fig.~\ref{fig:mscollapse}(b) must not be confused with a doping-independent quasiparticle residue. In the electron-doped system, states near the bottom of the upper Hubbard band reside predominantly on their minority sublattice, so their Fermi-level residue is governed by $\Zmin(0)$. This quantity varies strongly with doping and exceeds $\Rf$ substantially at strong coupling [orange versus blue curves in Fig.~\ref{fig:zall}(b)]. Because the two spin branches reside at different energies, a single quasiparticle residue at the Fermi level cannot characterize both branches simultaneously. Toward half filling, $\Zmin(0)$ smoothly connects to the local factor $\Zmin(E_{\rm bottom})$ at the band edge of the insulator, while $\Rf$ approaches its half-filled value. This behavior contrasts sharply with paramagnetic Mott physics, where the quasiparticle residue vanishes linearly with doping, $Z\propto n-1$ \cite{Werner2007}. In the altermagnet, the low-energy renormalization is set by the background magnetic order rather than by the carrier density, mirroring the behavior of the edge peak in the Bethe-lattice antiferromagnet~\cite{Sangiovanni2006}. Conversely, $\Zmin(0)$ decreases as doping increases, even as the carrier density grows, because doped carriers scatter off increasingly disordered local moments. At $U=15$, $\Zmin(0)$ approaches the paramagnetic value $Z=0.15$ at $n=1.15$.

\section{Spin-resolved spectral weight} \label{sec:spectra}
\begin{figure*}[t]
\includegraphics[width=\textwidth]{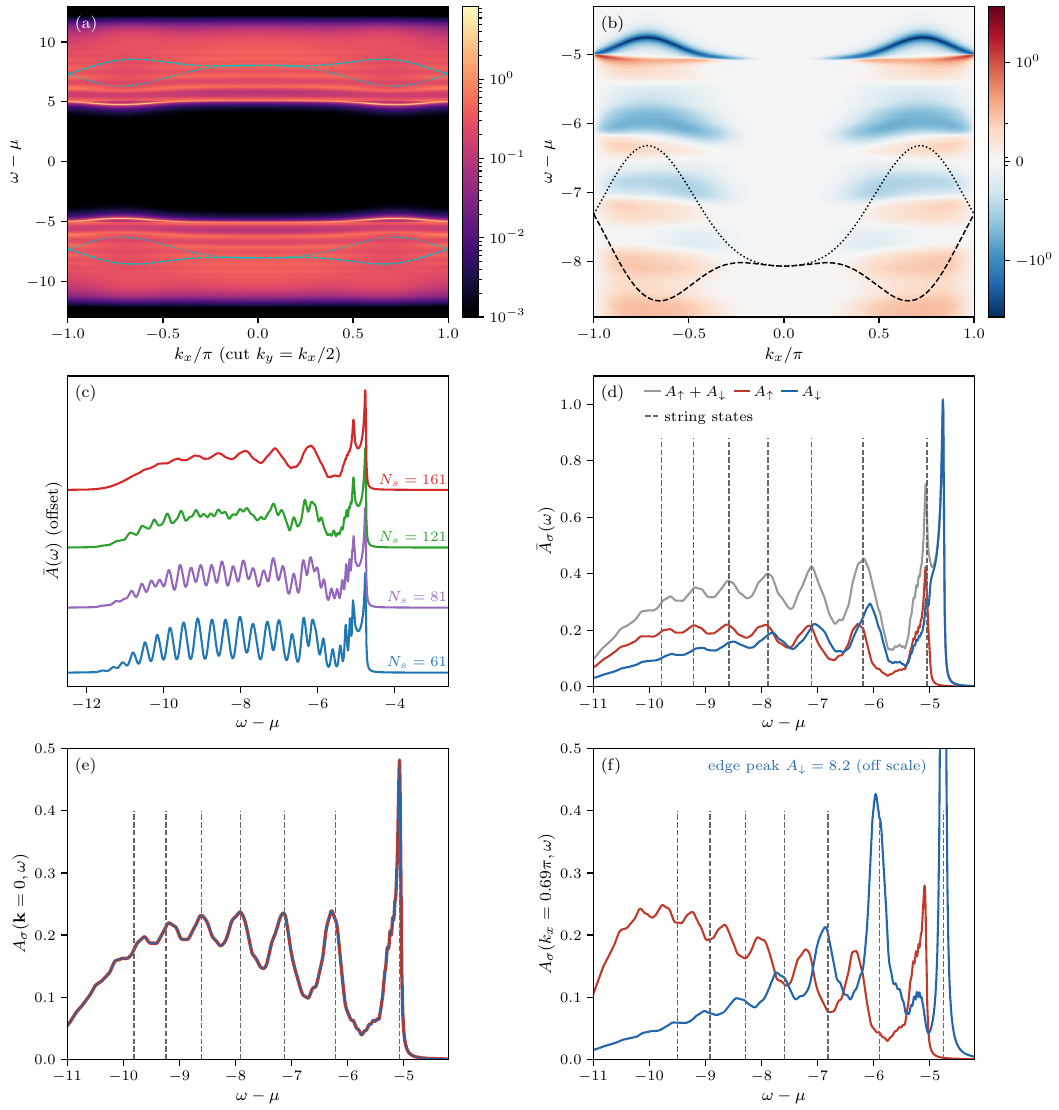}
\caption{Spectra of the half-filled altermagnet at $U=15$ along the cut $k_y=k_x/2$ from $T=0$ ED-DMFT. (a) $A_\uparrow+A_\downarrow$ for $N_s=161$ (truncated solver with $k=2$, logarithmic color scale) with the Hartree--Fock bands (dotted and dashed lines). (b) $A_\uparrow-A_\downarrow$ near the gap, on a symmetric logarithmic color scale. (c) Removal spectrum averaged along the cut for increasing bath size ($N_s=61$, $81$, $121$, and $161$, $k=2$), offset vertically. (d) Spin-resolved spectrum for $N_s=161$, averaged along the cut. Dashed lines: string states of the retraceable-path chain with $t_*=1.78$ (Appendix~\ref{app:strings}), measured from the edge-band peak of the spin sum. (e),(f) Spin-resolved spectra for $N_s=161$ at $k_x=0$, where the spins coincide, and at $k_x=0.69\pi$, the momentum of the maximal edge-band splitting (the spin-down edge peak, of height $8.2$, is cut off). Dashed lines: string states measured from the spin-down edge peak. All spectra are computed from the real-axis ED self-energy with a broadening $\eta=0.005$ in the lattice Green's function and are then smoothed in all panels with an energy-dependent Gaussian, which leaves the edge band sharp and removes the bath-discretization ripples of the sidebands. The standard deviation of the Gaussian is zero for $|\omega-\mu|\le5.0$, rises linearly to $0.03$, $0.05$, and $0.07$ at $|\omega-\mu|=5.6$, $6.5$, and $7.5$, and stays constant beyond.}
\label{fig:u15}
\end{figure*}

The suppression of the band splitting at strong coupling is accompanied by a pronounced redistribution of spectral weight. Figure~\ref{fig:u15} illustrates the half-filled spectrum at $U=15$. Although each Hubbard band spans roughly the bare bandwidth, only a small fraction of the total spectral weight resides in the coherent edge band nearest the gap [note the logarithmic intensity scale in Fig.~\ref{fig:u15}(a)]. Along the trajectory $k_y=k_x/2$, the spin-down branch forms a prominent dome at the top of the removal spectrum [Figs.~\ref{fig:u15}(b) and \ref{fig:u15}(f)], whereas the spin-up partner lies closer to the sidebands and is much weaker. Throughout this discussion, top and bottom denote the edges facing the insulating gap and the higher-binding-energy sidebands, respectively.

Averaging along the cut reveals an edge-band weight of $Z_{{\rm edge},\downarrow}=0.185$ for spin down compared to $Z_{{\rm edge},\uparrow}=0.055$ for spin up, corresponding to a large asymmetry ratio of $3.4$. Concurrently, their bandwidths renormalize to $0.18$ and $0.07$ of their Hartree--Fock counterparts. This spectral asymmetry grows across the crossover. The corresponding weight ratio is only $1.04$ at $U=3$ and $1.2$ at $U=4.5$ (Appendix~\ref{app:half}).

\begin{figure}[t]
\includegraphics[width=\columnwidth]{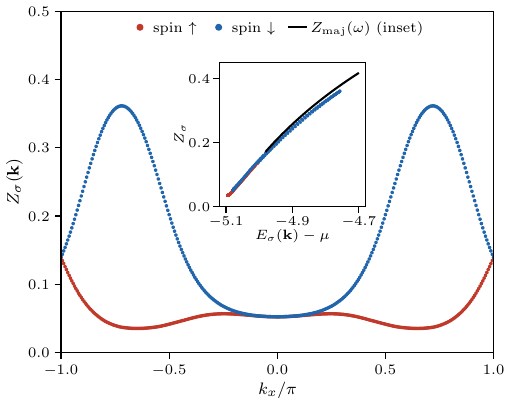}
\caption{Weight of the two removal edge-band branches at $U=15$ and $n=1$ along $k_y=k_x/2$ (ED, $N_s=161$, $k=2$), integrated from $0.1$ below the peak to the gap. Red: spin up, blue: spin down. Inset: the same weights plotted against the energy of each state. The black curve shows $\Zmaj(\omega)$ where $-\operatorname{Im}\Smaj<0.1$. Both branches approximately follow the same energy-dependent local factor, but the altermagnetic splitting places them at different energies. Near the top of the band, $Z_\downarrow$ lies slightly below $\Zmaj$, since the integration window misses about $6\%$ of the Lorentzian tail of the sharp spin-down peaks. The pole residues of Eq.~\eqref{eq:zspin} agree with $\Zmaj$ within $1\%$ there.}
\label{fig:zspin}
\end{figure}

To clarify the microscopic origin of this intensity imbalance, consider an isolated undamped pole at $E_\sigma(\bk)$ with normalized sublattice eigenvector $v$. Expanding the inverse lattice Green's function about this pole yields its residue via the projection of $1-\partial_\omega\operatorname{Re}\Sigma_\sigma$ onto $v$,
\begin{equation}
\frac{1}{Z_\sigma(\bk)}= \frac{|v_{\rm maj}|^2}{\Zmaj(E_\sigma)}+ \frac{|v_{\rm min}|^2}{\Zmin(E_\sigma)},
\label{eq:zspin}
\end{equation}
where the local renormalization factors $Z_\alpha$ [Eq.~\eqref{eq:Zalpha}] are evaluated at the pole energy. For broadened or overlapping features, we evaluate the integrated peak weight directly. The main panel of Fig.~\ref{fig:zspin} displays the resulting removal weights along the cut. While the weights coincide at $k_x=0$ and $\pm\pi$ due to symmetry-enforced degeneracy, the spin-down weight reaches $0.36$ near the momentum of maximal splitting, whereas the spin-up weight drops to $0.03$--$0.04$.

Equation~\eqref{eq:zspin} demonstrates that a local self-energy does not produce equal spectral weights for the two spin branches. At $U=15$, both removal branches reside predominantly on their respective majority sublattices, ensuring $Z_\sigma(\bk)\simeq\Zmaj[E_\sigma(\bk)]$. When plotted as a function of energy (inset of Fig.~\ref{fig:zspin}), the weights of both spin branches fall approximately onto a common curve. Near the top of the edge band, the majority self-energy is smooth, yielding $\Zmaj\simeq0.36$--$0.37$. Toward the bottom of the band, however, the states approach a self-energy pole of $\Smaj$, driving $\Zmaj$ below $0.05$. The altermagnetic splitting thus renders the strong energy dependence of a single local self-energy spin selective by positioning the two spin branches in disparate renormalization and damping regimes.

In the majority-sublattice limit, the energy profile of $\Zmaj$ dictates not only the branch intensities but also the magnitude of the band splitting. At fixed momentum, consider two removal poles at energies $E_1<E_2$, both polarized onto their majority sublattices. Their scalar pole conditions are $E_j+\mu-\operatorname{Re}\Smaj(E_j)=\epsilon_j$, with bare separation $\epsilon_2-\epsilon_1=2|\varepsilon_z|$. Subtracting these equations gives $E_2-E_1-[\operatorname{Re}\Smaj(E_2)-\operatorname{Re}\Smaj(E_1)]=2|\varepsilon_z|$. Provided $\Smaj$ contains no poles between the branches, the bracket equals the energy integral of $\partial_E\operatorname{Re}\Smaj$, and Eq.~\eqref{eq:Zalpha} yields
\begin{equation}
2|\varepsilon_z|=\int_{E_1}^{E_2}\frac{dE}{\Zmaj(E)} =\frac{E_2-E_1}{\bar Z_{\rm maj}},
\label{eq:window}
\end{equation}
where $\bar Z_{\rm maj}$ denotes the harmonic mean of $\Zmaj(E)$ across the inter-branch energy window. At the wavevector $\bk_*$ maximizing the splitting, $E_2-E_1=\DEc$, and Eq.~\eqref{eq:LR} gives
\begin{equation}
\Rf\simeq\frac{2|\varepsilon_z(\bk_*)|}{\Fc(\heff)}\,\bar Z_{\rm maj}.
\label{eq:Rwindow}
\end{equation}
While the individual peak weights sample $\Zmaj$ solely at the branch energies, the splitting is renormalized by the harmonic mean across the entire window. Consequently, $\Rf$ need not lie close to the individual edge-band weights. Equation~\eqref{eq:Rwindow} requires negligible sublattice mixing, weak damping, and no self-energy pole between the branches. At $U=15$, the lower branch at $\bk_*$ approaches poles of $\Smaj$, where $-\operatorname{Im}\Smaj$ reaches $0.6$, so Eq.~\eqref{eq:Rwindow} yields $0.21$ compared to the directly measured $\Rf=0.15$. The values of $\Rf$ in Figs.~\ref{fig:dmftU}, \ref{fig:zall}, and \ref{fig:mscollapse} are extracted directly from the physical spectral peaks via Eq.~\eqref{eq:LR}.

The spectral weight missing from the edge band is transferred to higher binding energies. Below the coherent edge, the ED spectra exhibit a dense manifold of discrete peaks that proliferate with increasing bath size [Fig.~\ref{fig:u15}(c)]. On top of this fine structure, a superstructure of broader sidebands emerges, at energies consistent with an effective retraceable-path string chain [dashed lines in Figs.~\ref{fig:u15}(d)--\ref{fig:u15}(f), Appendix~\ref{app:strings}]. This polaronic structure is particularly evident at $k_x=0$, where the two spins coincide and the dominant ladder peaks lie at the string-state energies [Fig.~\ref{fig:u15}(e)]. At the maximal splitting, the first spin-down sideband replicates the edge dispersion [Fig.~\ref{fig:u15}(f)]. The sidebands retain a clear spin polarization, with spin up dominating the deep lower Hubbard band [Fig.~\ref{fig:u15}(d)]. While same-sublattice hopping permits hole motion without generating nearest-neighbor spin defects, the effective string picture captures the sideband spacing within about $10\%$ (Appendix~\ref{app:strings}).

Spectral moments offer a complementary perspective. The first moment of the $2\times2$ spectral matrix follows from the high-frequency expansion of the lattice Green's function $G_\sigma(\bk,z)\sim1/z+(h(\bk)-\mu+U\operatorname{diag}(n_{A\bar\sigma},n_{B\bar\sigma}))/z^2$:
\begin{equation}
\int d\omega\,\omega A_\sigma(\bk,\omega) =h(\bk)-\mu+U\operatorname{diag}(n_{A\bar\sigma},n_{B\bar\sigma}).
\label{eq:moment}
\end{equation}
The right-hand side is the static Hartree--Fock matrix evaluated using the fully interacting densities. At half filling, its eigenvalues are $\pm R_\sigma(\bk)$, representing the Hartree--Fock dispersion evaluated with the exact DMFT magnetization. We define the centroid of the occupied removal spectrum as
\begin{equation}
\bar\omega_\sigma(\bk)=\frac{\int_{\omega\le0}\omega\,\operatorname{Tr}A_\sigma(\bk,\omega)\,d\omega}{\int_{\omega\le0}\operatorname{Tr}A_\sigma(\bk,\omega)\,d\omega},
\label{eq:centroid}
\end{equation}
where the trace sums over sublattices. Equation~\eqref{eq:moment} ensures $\bar\omega_\sigma(\bk)=-R_\sigma(\bk)$ whenever the removal spectrum at $\bk$ contains the full weight of the lower Hartree--Fock eigenvector and none of the weight of the upper one. Deviations between $\bar\omega_\sigma$ and $-R_\sigma$ thus measure how much the sublattice composition of the removal states differs from that of the lower Hartree--Fock band.

\begin{figure}[t]
\includegraphics[width=\columnwidth]{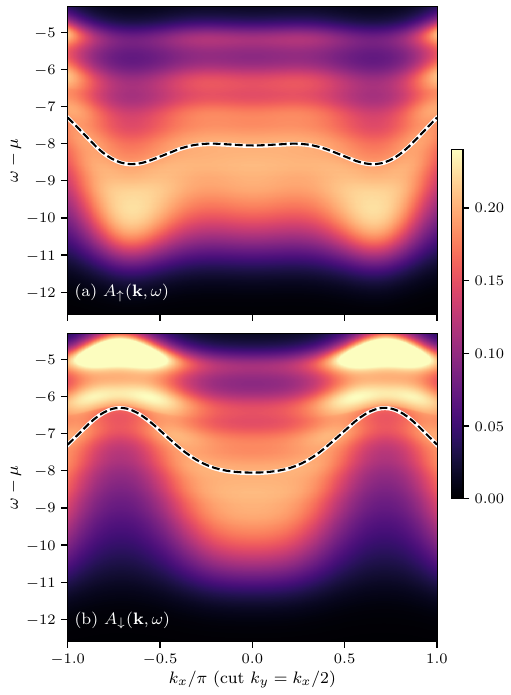}
\caption{Removal spectra at $U=15$, $n=1$ ($N_s=161$, $k=2$), smoothed with a Lorentzian of width $0.35$. (a) Spin up and (b) spin down, with a common color scale. White lines show the centroids of the unsmoothed removal spectra, and black dashed lines the Hartree--Fock dispersions. 
}
\label{fig:envelope}
\end{figure}

At $U=15$, the removal spectrum carries $99.9\%$ of the weight of one band for each spin projection and wavevector, and the DMFT and Hartree--Fock moments differ by less than $0.3\%$.
As a result, the removal centroids track the Hartree--Fock bands across the entire cut to within $0.011$ (Fig.~\ref{fig:envelope}), despite the fact that spectral weight is spread across a sideband ladder spanning from $-4.8$ to $-11.5$. In DMFT, the spectral centroids in the strong-correlation regime thus preserve the unrenormalized Hartree--Fock $g$-wave splitting, whereas the coherent edge-band splitting is strongly reduced. Physically, the first moment measures the instantaneous energy of the bare hole state $c_{\bk\sigma}|\Psi_0\rangle$ created in the ground state $|\Psi_0\rangle$, and is linked to the short-time propagation. In the Mott state, the exchange field locks each spin to a single sublattice, and the spin dependence of this short-time energy is governed entirely by the same-sublattice hopping $\varepsilon_z$, which displaces the hole without disrupting the ordered spin background. The centroid splitting therefore approaches $2|\varepsilon_z|$. Conversely, intersublattice hopping generates spin defects and distributes spectral weight into the polaronic ladder. The long-time coherent motion of the dressed quasiparticle defines the edge band, whose splitting is heavily renormalized by $\Rf$.

This close match between the spectral centroids and Hartree--Fock dispersions is unique to the Mott regime. At weak interactions, the removal states do not have the same sublattice composition as the lower Hartree--Fock band, because the exchange field acting on them is frequency dependent ($\operatorname{Re}\delta\Sigma(\omega)$ instead of $\hex$). Projected onto the Hartree--Fock eigenvectors, about $3\%$ of the removal weight lies along the upper-band eigenvector, mostly at high binding energy, and correspondingly $3\%$ of the lower-band eigenvector appears in high-energy addition features. This shifts the cut-averaged centroids below $-R_\sigma$ by $0.15$--$0.21$ at $U=3$ and $0.23$--$0.34$ at $U=4.5$ [Figs.~\ref{fig:slater}(i)--\ref{fig:slater}(l)]. A detailed analysis of the Slater regime is provided in Appendix~\ref{app:half}.

\section{Summary} \label{sec:discussion}
Our results demonstrate that the staggered magnetic moment, low-energy quasiparticle splittings, polaronic sidebands, and spectral centroids probe different physical aspects of the altermagnetic state.
We also revealed two mechanisms through which local dynamical correlations alter the relationship between altermagnetic order and electronic excitations. First, at low temperatures, increasing the interaction strength amplifies both the staggered magnetization and spin--sublattice locking, but simultaneously enhances the dynamical self-energy dressing that narrows the coherent edge band. The competition between these two effects produces a maximal spin splitting at intermediate interaction ($U\approx W_0$). Second, in the strong-coupling regime, carrier doping rapidly screens the low-frequency exchange field encoded in the real part of the self-energy, pulling the locking factor $\Lf$ out of saturation. As a consequence, the spin splitting decreases strongly with doping, roughly in proportion to the ordered moment, contrary to static mean-field predictions of a robust, saturated splitting. 

The spin-resolved spectral functions directly mirror these many-body effects and undergo a qualitative change across the Slater--Mott crossover. In the weakly correlated Slater insulator, the edge bands are the only coherent features of the removal spectrum. In the strongly correlated Mott insulator, most of the removal weight is transferred into a ladder of polaronic sidebands, leaving two highly asymmetric edge branches. The weight asymmetry in the edge states arises because the altermagnetic splitting places the two spin projections in different self-energy regimes near the Mott gap edge. These results are consistent with the spin-selective weight transfer found in recent strong-coupling studies~\cite{Daghofer2026,Lanzini2026} and complement the dynamical reconstruction explored in Ref.~\cite{DelRe2025}. Remarkably, while the coherent edge splitting is heavily suppressed, the first moments (centroids) of the removal spectra track the unrenormalized Hartree--Fock dispersions, and hence exhibit a much larger altermagnetic splitting.  

Our conclusions are based on the symmetry-constrained collinear solution of single-site DMFT. This approach captures local, sublattice-resolved dynamical dressing, while omitting nonlocal self-energy corrections from dispersing spin waves and string-relaxation dynamics. Spatial fluctuations in two dimensions can further dress polaron dispersions and broaden the weaker spin branch into an incoherent continuum~\cite{Kane1989,Martinez1991,Daghofer2026,Lanzini2026}. Cluster or diagrammatic extensions of DMFT~\cite{Maier2005,Rohringer2018} would make it possible to assess the impact of nonlocal correlations. Furthermore, at nonzero temperature, our Hartree--Fock and single-site DMFT treatments describe a mean-field magnetic background, representative of weakly coupled layers rather than a strictly two-dimensional system.

\begin{acknowledgments}
The DMFT calculations were performed on the Beo05 cluster at the University of Fribourg. S.A.A.G was supported by the U.S. Department of Energy, Office of Science, Basic Energy Sciences, under Award No. DE-SC0024942. P.W. and S.A.A.G. acknowledge the hospitality of the Aspen Center for Physics, where this project was started during the 2026 summer program. The Aspen Center for Physics is supported by the National Science Foundation under Grant No.~PHY-2210452 and by a grant from the Simons Foundation (1161654, Troyer).

\textit{Use of AI tools.} The Hartree--Fock code and the exact-diagonalization DMFT code, including the truncated configuration-interaction solver, were implemented with the assistance of Anthropic's Claude Opus 5.5. The DMFT results were benchmarked against published data and against our own CT-QMC code. Claude Opus 5.5 was also used for the data analysis, the plotting of the figures, and the editing of the manuscript. The AI tool was directed through iterative prompting, and all results were checked by the authors, who take full responsibility for the content of this paper.
\end{acknowledgments}

\begin{figure}[b]
\includegraphics[width=\columnwidth]{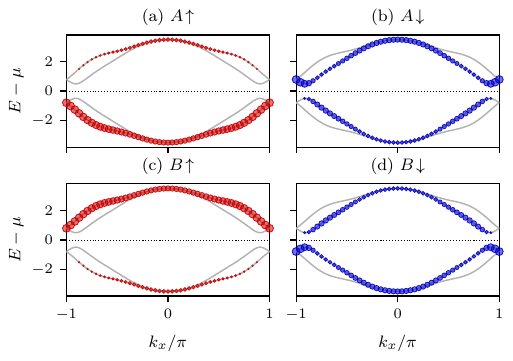}
\caption{Sublattice- and spin-resolved Hartree--Fock bands along $k_y=k_x/2$ at $U=3$, $n=1$. The symbol size is proportional to the weight of the component indicated in each panel. Gray lines show all four bands, and the dotted line marks the chemical potential.}
\label{fig:fat}
\end{figure}

\begin{figure*}[t]
\includegraphics[width=\textwidth]{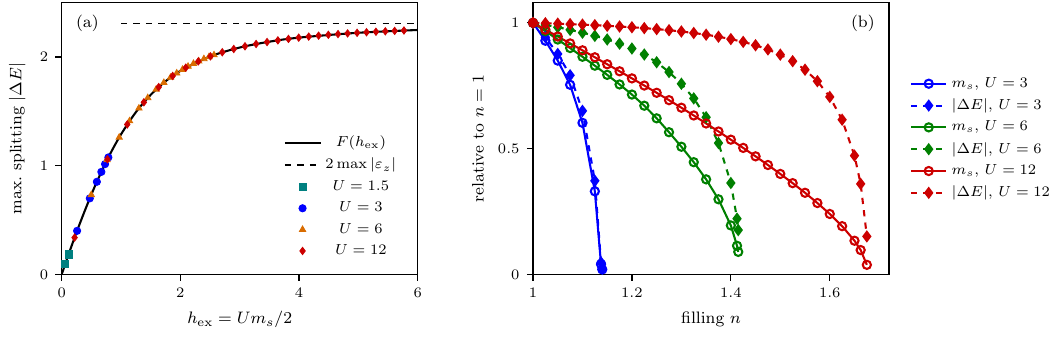}
\caption{Hartree--Fock locking reference at $T=0.01$. (a) Full-zone maximum splitting versus $\hex=Um_s/2$ for $U=1.5$, $3$, $6$, and $12$, with filling varied from half filling to the boundary of the ordered state. The line is $F(\hex)$. (b) Moment (circles) and splitting (diamonds) versus filling at $U=3$, $6$, and $12$, normalized to their half-filled values.}
\label{fig:collapse}
\end{figure*}

\appendix
\section{Hartree--Fock reference} \label{app:hf}
\subsection{Band character, locking, and doping}
Expanding the signed Hartree--Fock energy splitting $\Delta E_\nu=\nu(R_\uparrow-R_\downarrow)$ in powers of the exchange field yields
\begin{equation}
\Delta E_\nu\simeq
\begin{cases}
-2\nu\hex\varepsilon_z/(\varepsilon_x^2+\varepsilon_z^2)^{1/2},&\hex\to0,\\[2pt]
-2\nu\varepsilon_z[1-\varepsilon_x^2/(2\hex^2)],&\hex\gg|\varepsilon_x|,|\varepsilon_z|.
\end{cases}
\label{eq:limits}
\end{equation}
The small-field expansion holds away from the zone boundary, where $\varepsilon_x=\varepsilon_z=0$ pins the splitting to zero for all fields. In the strong-field limit, an intersublattice hop transfers an electron to its minority sublattice at an energy cost $2\hex$, contributing only virtual corrections. To leading order, the two lower branches reduce to the opposite same-sublattice dispersions. 

The sublattice weight on site $A$ is given by $[1+\nu(\varepsilon_z-\sigma\hex)/R_\sigma]/2$. At $U=3$, the lower spin-up branch exhibits $61\%$ $A$-sublattice character near the $\Gamma$ point where intersublattice mixing is maximal, becoming fully sublattice polarized at the zone boundary (Fig.~\ref{fig:fat}). Averaged along the cut $k_y=k_x/2$, the $A$-sublattice weight is $79\%$. This wave-function polarization, rather than the bare magnetic moment, governs the splitting. As $\hex$ increases, intersublattice mixing is suppressed and the splitting approaches $2|\varepsilon_z|$ [Eq.~\eqref{eq:limits}]. Once the bands are completely polarized onto single sublattices, further moment growth cannot increase the splitting.

The maximum splitting across the full Brillouin zone, $F(\hex)$, and the cut maximum, $\Fc(\hex)$, differ little, with saturation ceilings of $2.31$ and $2.26$ and half-saturation fields $h_{1/2}=0.86$ and $0.85$, respectively. In Fig.~\ref{fig:collapse}(a), the Hartree--Fock solutions for various interactions and dopings lie on $F(\hex)$, as Eq.~\eqref{eq:F} requires. At large $U$, carrier doping primarily removes excess locking, leaving the splitting pinned near saturation. For example, at $U=12$, the moment decreases from $0.961$ to $0.515$ between $n=1$ and $1.4$, while the full-zone splitting changes only from $2.244$ to $2.099$. At $U=3$, the field is already below half saturation at half filling, so doping to $n=1.1$ reduces the moment and splitting by $40\%$ and $35\%$, respectively. Thus, the weak-coupling doping response is qualitatively consistent between Hartree--Fock and DMFT, whereas the strong-coupling response differs fundamentally.

\subsection{Collinear phase competition and spiral order} \label{sec:phase}
\begin{figure}[t]
\includegraphics[width=\columnwidth]{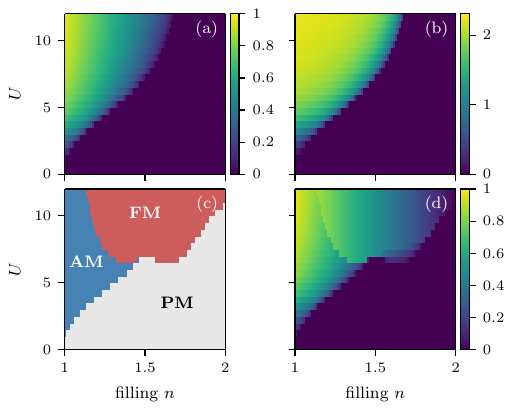}
\caption{Collinear Hartree--Fock phase diagram at $T=0.01$. (a) Staggered moment and (b) maximum upper-band splitting within the altermagnetic solution. (c) Lowest-free-energy state among the altermagnetic (AM), ferromagnetic (FM), and paramagnetic (PM) solutions. (d) Moment of that state.}
\label{fig:phase}
\end{figure}

Within the altermagnetic Hartree--Fock channel, the ordered region in the interaction--filling plane expands with $U$ [Fig.~\ref{fig:phase}(a)]. At half filling and $T=0.01$, order appears between $U=0.75$ and $1$, and the moment approaches unity at large $U$. The large-$U$ extent of this mean-field ordered region is not compatible with the superexchange scale.
The physical nearest-neighbor hopping between $A$ and $B$ sites in Eq.~\eqref{eq:eps} is $t_x/4$, yielding the antiferromagnetic superexchange
\begin{equation}
J_{AB}=\frac{4(t_x/4)^2}{U}=\frac{t_x^2}{4U}.
\label{eq:J}
\end{equation}
While $J_{AB}$ scales as $1/U$, the Hartree--Fock exchange field $Um_s/2$ grows linearly with $U$.

A uniform ferromagnetic state has order parameter $m_f\equiv m_A=m_B$ and produces a momentum-independent Zeeman splitting $Um_f$. Comparing the free energies at fixed filling of the altermagnetic, ferromagnetic, and paramagnetic solutions reveals that the ferromagnet emerges between $U=6$ and $6.5$, occupying an extensive portion of the doped phase diagram [Fig.~\ref{fig:phase}(c)]. It is fully polarized ($m_f=2-n$) near the altermagnetic boundary and across the entire region for $U\gtrsim8.5$, with a first-order phase transition separating the two magnetic states. Because Hartree--Fock is known to severely overestimate ferromagnetism in doped Hubbard models~\cite{Kanamori1963,Hirsch1985,Shastry1990,Vollhardt1997}, we restrict the investigation in the main text to the altermagnetic and paramagnetic states.

The stability against noncollinear spiral order can be assessed in the strong-coupling Heisenberg limit at half filling. In addition to the intersublattice exchange $J_{AB}=t_x^2/(4U)$ [Eq.~\eqref{eq:J}], the same-sublattice hoppings $t_1$ and $t_z$ generate antiferromagnetic exchanges $J_1=t_1^2/U$ and $J_g=t_z^2/(16U)$, respectively. These same-sublattice terms frustrate the intersublattice exchange that stabilizes the collinear N\'eel state. For a planar spin spiral with pitch vector $\mathbf Q$, expanding the lowest eigenvalue of the exchange matrix, $\lambda_-(\mathbf Q)$, yields
\begin{equation}
\lambda_-(\mathbf Q)-\lambda_-(0)=\frac{t_x^2-8t_1^2-5t_z^2}{8U}|\mathbf Q|^2+O(Q^4).
\label{eq:stiffness}
\end{equation}
The collinear altermagnetic ground state ($\mathbf Q=0$) is locally stable against long-wavelength spiral distortions provided $t_x^2>8t_1^2+5t_z^2$.

For our model parameters ($t_1=0$, $t_x=3.4$, and $t_z=1.5$), the spin stiffness is small but positive, $t_x^2-5t_z^2=0.31$, so that the collinear state is stable against long-wavelength spirals. Furthermore, $\mathbf Q=0$ constitutes the global minimum of $\lambda_-(\mathbf Q)$ across the full Brillouin zone, establishing stability against all planar spirals in the Heisenberg limit. This result does not establish stability at finite interaction or doping, where mobile carriers could favor incommensurate order or phase separation. Our Hartree--Fock and DMFT calculations therefore characterize the collinear solutions without establishing their stability against these states.

\section{Weak-coupling spectra and band weights} \label{app:half}
\begin{figure*}[t]
\includegraphics[width=\textwidth]{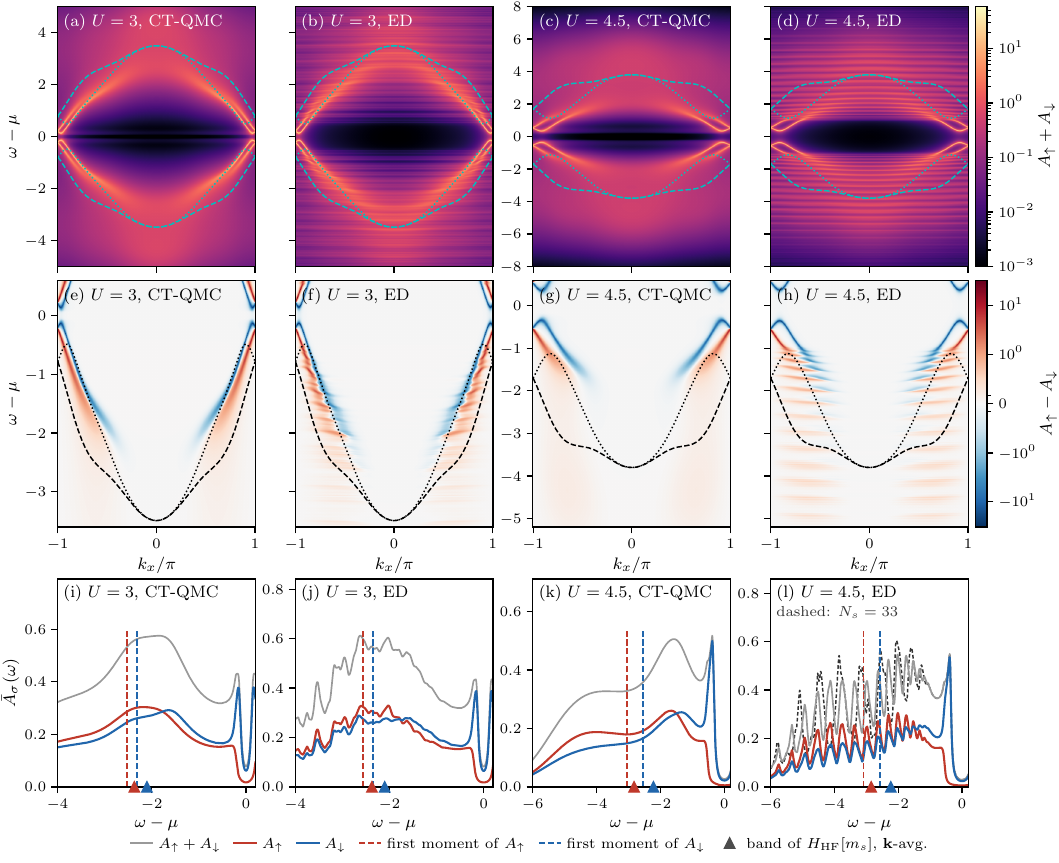}
\caption{Spectra in the Slater regime at half filling along the cut $k_y=k_x/2$: CT-QMC at $T=0.01$ and ED at $T=0$ for $U=3$ [$N_s=33$, $k=4$, panels (a),(b),(e),(f),(i),(j)] and $U=4.5$ [$N_s=61$, $k=3$, panels (c),(d),(g),(h),(k),(l)]. Top row: $A_\uparrow+A_\downarrow$ on a logarithmic color scale, with the Hartree--Fock bands (dashed: spin up, dotted: spin down). Second row: $A_\uparrow-A_\downarrow$ on the removal side, on a symmetric logarithmic color scale, with the Hartree--Fock bands. Each of these two rows uses a common color scale. Third row: spin-resolved removal spectra averaged along the cut, broadened by $0.03$ (spin up red, spin down blue, sum gray). Dashed vertical lines: first moments of $A_\uparrow$ and $A_\downarrow$ over $\omega\le0$. Triangles: band energies of the Hartree--Fock Hamiltonian with the exact moment, averaged along the cut. Dashed gray line in (l): sum for $N_s=33$ ($k=4$).}
\label{fig:slater}
\end{figure*}

At $U=3$ and $4.5$, half-filled CT-QMC simulations enable direct benchmarking against ED results (Fig.~\ref{fig:slater}). Maximum-entropy analytical continuation of the CT-QMC data yields continuous dispersive Slater bands that are moderately renormalized relative to Hartree--Fock, with a small zone-boundary gap and a maximum splitting near $|k_x|\approx0.9\pi$. The two techniques agree almost quantitatively. CT-QMC (ED) yields moments $m_s=0.279$ ($0.280$) and $0.536$ ($0.547$), cut gaps of $0.27$ ($0.28$) and $0.70$ ($0.73$), maximal splittings of $0.30$ ($0.28$) and $0.58$ ($0.55$), and renormalization factors $\Rf=0.73$ ($0.70$) and $0.55$ ($0.52$) at $U=3$ and $4.5$, respectively. Both methods yield an identical exchange-field fraction $\heff/(Um_s/2)=0.64$. Because finite ED baths cannot fully resolve smooth continuous bands, dispersion-dependent quantities (such as band weights and widths) are extracted from CT-QMC, while the ED splitting is evaluated at the momentum of the CT-QMC maximum ($k_x=0.91\pi$ and $0.87\pi$). In Fig.~\ref{fig:dmftU}, the half-filled data for $U\le4.5$ are computed with the CT-QMC solver.

At $U=3$, the ED removal spectrum consists of the dispersive spin-split band, discretized into steps by the finite bath. At $U=4.5$, additional nondispersive weight emerges below the coherent band across the entire cut [Fig.~\ref{fig:slater}(h)]. The spacing of these features becomes finer as the bath size increases [Fig.~\ref{fig:slater}(l)]. Since this structure is not converged at the largest accessible $N_s$, we interpret it as a discretized representation of a smooth continuum, as found by CT-QMC.

\begin{figure}[t]
\includegraphics[width=\columnwidth]{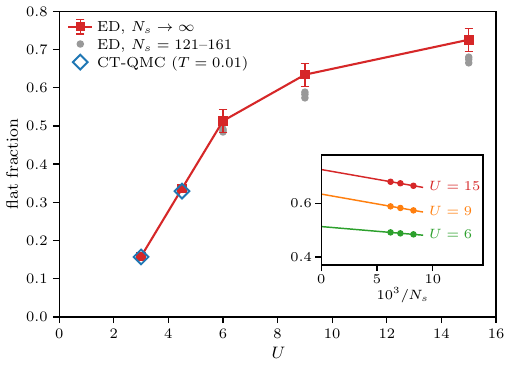}
\caption{Flat fraction of the removal weight at half filling. Squares: ED, for $U=3$ and $4.5$ with the largest bath ($N_s=33$, $k=4$ and $N_s=61$, $k=3$), for $U\ge6$ extrapolated to $N_s\to\infty$ (error bars: estimated uncertainty of the extrapolation). Diamonds: CT-QMC results. Inset: flat fraction versus $1/N_s$ with linear fits. Gray dots: ED results for $N_s=121$--$161$ ($k=2$) used in the fits.}
\label{fig:flat}
\end{figure}

To quantify this momentum-independent weight in a way that is relatively insensitive to the discretization, we define the \textit{flat fraction} as the portion of removal weight shared by all momenta along the cut, $\int\min_{\bk}A(\bk,\omega)\,d\omega\big/\int\langle A(\bk,\omega)\rangle_{\bk}\,d\omega$ for $\omega\le0$ (broadened by $0.03$). The flat fraction rises sharply from $U=3$ to $U=6$, leveling off at stronger coupling (Fig.~\ref{fig:flat}). At $U=3$ and $4.5$, ED and CT-QMC agree, and the ED values are converged with bath size. For $U\ge6$, the flat fraction continues to grow with $N_s$ as the bath resolves the sideband ladder, extrapolating linearly in $1/N_s$ to the infinite-bath limit (inset of Fig.~\ref{fig:flat}). The momentum-independent weight at $U=4.5$ represents the precursor continuum of the polaronic ladder, which develops into distinct sidebands at strong coupling. At $U=3$, the small flat fraction mainly reflects the broadening of the dispersive bands.

The momentum-averaged weak-coupling spectra [Figs.~\ref{fig:slater}(i)--\ref{fig:slater}(l)] also exhibit the spin asymmetry discussed in Sec.~\ref{sec:spectra}. At both interactions, the edge peak nearest the gap is dominated by spin down, while spin up has more weight at higher binding energy. Consequently, the first moment of $A_\uparrow$ resides below that of $A_\downarrow$ on the removal side. Unlike for the Mott state at $U=15$, the removal centroids (dashed lines) however deviate substantially from the Hartree--Fock eigenvalues (triangles), by $0.15$--$0.34$, because of the high-energy weight discussed at the end of Sec.~\ref{sec:spectra}. CT-QMC and ED match on these first moments to within $0.04$.

The spin-resolved weights and bandwidth ratios for the edge band are summarized in Table~\ref{tab:zspin}. The integrated weight traces the peak from $0.1$ below the maximum to the gap, matching the pole residue for isolated undamped peaks. In the Slater regime, the band weights $Z_{{\rm band},\sigma}$ for the two spin projections differ by only $4\%$ at $U=3$ and $20\%$ at $U=4.5$, displaying comparable bandwidth ratios ($0.76$ and $0.74$ at $U=3$, and $0.71$ and $0.67$ at $U=4.5$). This near-degeneracy reflects the fact that local self-energy factors remain weakly spin dependent while the wave functions retain strong intersublattice mixing.

\begin{table}[t]
\caption{Band weights and widths at half filling along $k_y=k_x/2$. $U=3$ and $4.5$ refer to the CT-QMC band nearest the gap, and $U=9$ and $15$ to the ED edge band ($N_s=61$, $k=3$, and $N_s=161$, $k=2$). For $U=9$, $N_s=161$ ($k=2$) gives the same weights within $0.002$ and width ratios within $0.01$. The last two columns list the width ratios for spin up and spin down.}
\label{tab:zspin}
\begin{ruledtabular}
\begin{tabular}{cccccc}
$U$ & $Z_{{\rm band},\uparrow}$ & $Z_{{\rm band},\downarrow}$ & $Z_{{\rm band},\downarrow}/Z_{{\rm band},\uparrow}$ & $W_\uparrow/W_\uparrow^{\rm HF}$ & $W_\downarrow/W_\downarrow^{\rm HF}$ \\ \hline
3 & 0.79 & 0.83 & 1.04 & 0.76 & 0.74\\
4.5 & 0.51 & 0.62 & 1.2 & 0.71 & 0.67\\
9 & 0.077 & 0.252 & 3.3 & 0.16 & 0.26\\
15 & 0.055 & 0.185 & 3.4 & 0.07 & 0.18\\
\end{tabular}
\end{ruledtabular}
\end{table}

\section{Effective string-chain model} \label{app:strings}
In an altermagnetic Mott insulator, intersublattice hopping of a hole generates a string of misaligned spins. Two approximations render this many-body problem analytically tractable. In the Ising limit, transverse spin fluctuations are neglected, preventing the self-healing of spin defects and causing the string potential to grow linearly with length. In the retraceable-path approximation, the hole can return to the origin only by exactly retracing its trajectory, excluding closed loops~\cite{Trugman1988}. Both approximations become exact on the Bethe lattice in the limit of infinite coordination, where closed loops are absent and transverse spin fluctuations are suppressed~\cite{Brinkman1970,Strack1992}. This is also the limit in which single-site DMFT becomes exact. For the Bethe lattice problem, the single-particle Green's function for the hole satisfies the continued fraction
\begin{equation}
G(\omega)=\frac{1}{\omega-t_*^2G(\omega-V)},
\label{eq:chain}
\end{equation}
corresponding to the edge Green's function of a semi-infinite tight-binding chain of string lengths $\ell=0,1,\ldots$, with hopping amplitude $t_*$ and linear potential $\ell V$. Here, $V$ represents the exchange cost of extending the string by one lattice step, set by the superexchange $J_{AB}$ [Eq.~\eqref{eq:J}]. The Bethe lattice with infinite coordination has a semicircular density of states, which differs from that of our lattice model, but we can use the Bethe-lattice results by matching the second moment of the density of states. Because the potential increases linearly with $\ell$, all eigenstates on the chain are localized, producing a discrete ladder of $\delta$-function excitations. In the weak-confinement regime ($V\ll t_*$), the excitation energies above the lowest state are $\Delta E_n=(a_n-a_1)t_*^{1/3}V^{2/3}$, where $a_n$ are the magnitudes of the zeros of the Airy function.

\begin{figure}[t]
\includegraphics[width=\columnwidth]{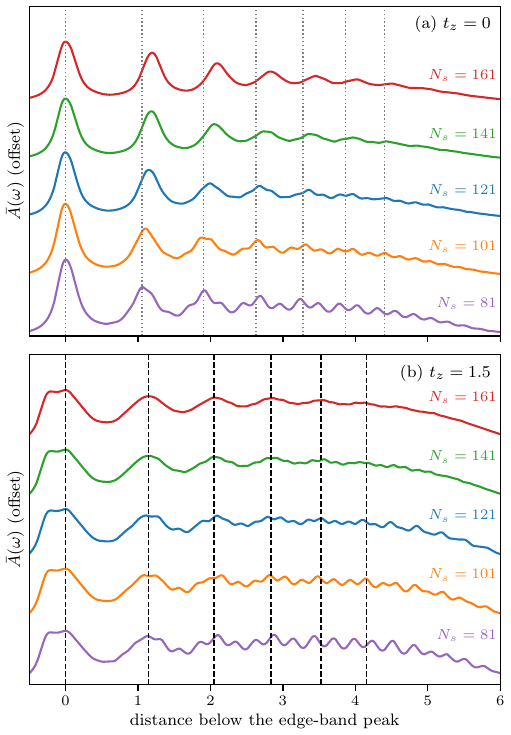}
\caption{Cut-averaged removal spectra at $U=15$ for $N_s=81$, $101$, $121$, $141$, and $161$ ($k=2$). (a) Conventional antiferromagnet ($t_z=0$) and (b) altermagnet ($t_z=1.5$). Spectra are aligned at the edge peak, offset vertically, and broadened by $0.1$. Each curve is normalized by its largest value more than $0.3$ below the edge peak. Dotted lines in (a): chain excitations for $t_*=1.70$. Dashed lines in (b): chain excitations for $t_*=1.78$. Both chain calculations use $V=2t_*^2/U$.}
\label{fig:afm}
\end{figure}

Setting $t_z=0$ recovers the conventional N\'eel antiferromagnet, with second moment $t_*^2=\langle\varepsilon_x^2\rangle=t_x^2/4$, giving $t_*=1.70$. On a lattice with coordination number $z$ and hopping $t_*/\sqrt{z}$, an additional string segment increases the number of frustrated bonds by $z-2$, each costing $J/2$ with $J=4t_*^2/(zU)$, so that $V=(z-2)J/2=(1-2/z)\,2t_*^2/U$. Matching the second moment fixes $t_*$, and hence $zJ$, but not this bond counting. In the infinite-coordination limit underlying single-site DMFT, the string-step potential is $V=2t_*^2/U$. For $z=4$, where $t_*/\sqrt{z}=t_x/4$ is the actual nearest-neighbor hopping and $J=J_{AB}$ [Eq.~\eqref{eq:J}], the same counting gives only $V=J_{AB}=t_*^2/U$, half the DMFT value.
At $U=15$, $V=0.385$, and direct diagonalization of the semi-infinite chain predicts excitations at $1.06$, $1.90$, $2.63$, and $3.28$ above the ground state, corresponding to removal sidebands below the edge peak. The Airy expression instead gives $1.11$, $2.01$, $2.81$, and $3.54$. It is an asymptotic formula for weak confinement ($V\ll t_*$), in which the lowest states extend over many chain sites, whereas here the confinement length $(t_*/V)^{1/3}\simeq1.64$ spans only a few chain sites.
Figure~\ref{fig:afm}(a) compares these chain excitations with the ED spectra of the antiferromagnet ($t_z=0$) for $N_s=81$--$161$. Without free parameters, the sideband maxima (extracted at broadening $0.2$) reside at $1.20$, $2.10$, $2.84$, and $3.45$ below the edge peak at $N_s=161$, exceeding the chain predictions by only $0.14$--$0.21$ ($5$--$13\%$), consistent with $O(t_*/U)$ corrections. While the edge-band weight is fully converged ($\Zedge=0.20$ for $N_s=141$ and $161$), the sideband peaks still drift outward by $0.04$--$0.15$ between $N_s=121$ and $161$.

For the altermagnet with both intersublattice and same-sublattice hopping, the second moment of the noninteracting dispersion is
\begin{equation}
t_*^2=\langle\varepsilon_x^2+\varepsilon_z^2\rangle =\frac{t_x^2}{4}+\frac{t_z^2}{8},
\label{eq:tstar}
\end{equation}
which for $t_x=3.4$ and $t_z=1.5$ yields $t_*=1.78$. Adopting the same effective parametrization $V=2t_*^2/U$ produces chain levels at $1.14$, $2.05$, $2.83$, and $3.53$. As shown in Fig.~\ref{fig:afm}(b), the sideband maxima in the altermagnet align with these chain predictions within $0.05$ across $N_s=121$--$161$, which supports their interpretation as string states. The altermagnetic sidebands are broader than their antiferromagnetic counterparts because of the spin splitting and because same-sublattice hopping permits coherent hole motion without creating spin defects.

The chain model also gives some hint why $\Rf$ decreases approximately as $1/U$. Same-sublattice hopping moves the hole without creating spin defects only if no string is attached. Otherwise, the hole detaches from the string, leaving the retraceable-path space, and the spin that fills the vacated site is frustrated by the misaligned spin at the end of the string. If $\varepsilon_z$ acts only on the string-free state $\ell=0$, the chain Hamiltonian acquires the term $\pm\varepsilon_z|0\rangle\langle0|$ for the two spins, and the Hellmann--Feynman theorem gives $dE_0/d\varepsilon_z=\pm P_0$, where $E_0$ is the lowest chain level and $P_0=|\psi_0|^2$ is the zero-length-string probability, which is also the weight of the lowest pole. To linear order in $\varepsilon_z$, the splitting is $2P_0|\varepsilon_z|$ instead of $2|\varepsilon_z|$, so it is reduced by the same factor as the pole weight. In the weak-confinement limit, $P_0\simeq V/t_*=2t_*/U$, which motivates the approximate $1/U$ dependence of $\Rf$. For sufficiently large $\varepsilon_z$, the two branches lie at different energies. Their weights are $P_0$ at these energies, whereas the splitting, $\int_{-|\varepsilon_z|}^{|\varepsilon_z|}P_0(\varepsilon)\,d\varepsilon$, averages $P_0$ over the window between them, analogous to $\bar Z_{\rm maj}$ in Eq.~\eqref{eq:Rwindow}. With $t_*=1.78$ and $|\varepsilon_z|=1.13$ at $\bk_*$, the two branches have $P_0=0.53$ and $0.16$ at $U=9$ and $P_0=0.42$ and $0.09$ at $U=15$, which qualitatively reproduces the spin asymmetry of the edge-band weights. A related weighting by the zero-length-string probability appears in the parton theory of Ref.~\cite{Lanzini2026}.

The polaronic sidebands also inherit the altermagnetic spin splitting. Table~\ref{tab:sideband} lists the spin-resolved peak energies at $k_x=0.69\pi$, the momentum of maximal edge-band splitting [Fig.~\ref{fig:u15}(f)]. Each spin projection supports an independent ladder of sidebands shifted nearly rigidly by the edge splitting. Relative to their respective edge peaks, the first three sidebands lie at $1.19$, $2.13$, and $2.94$ for spin down, and $1.24$, $2.15$, and $2.94$ for spin up, close to the chain values $1.14$, $2.05$, and $2.83$. Concurrently, spectral weight is redistributed across the lower Hubbard band, with spin down dominating the upper edge and spin up dominating the deep interior [Fig.~\ref{fig:u15}(b)]. At $k_x=0.69\pi$, the weight below $\omega-\mu=-9$ is $0.46$ for spin up compared to only $0.10$ for spin down.

\begin{table}[t]
\caption{Spin-resolved peak energies $\omega-\mu$ of the edge band and the first three sidebands at $U=15$, $n=1$, and $k_x=0.69\pi$, $k_y=k_x/2$ ($N_s=161$, $k=2$). The sideband peaks are taken from spectra with $\eta=0.02$ and an additional Gaussian broadening of $0.1$. Their positions change by less than $0.1$ for $N_s=121$--$161$ and broadenings $0.05$--$0.2$. The spin-up edge peak consists of several damped poles, and its energy is taken from the band tracking used in Fig.~\ref{fig:zspin}.}
\label{tab:sideband}
\begin{ruledtabular}
\begin{tabular}{lcccc}
& edge & 1st & 2nd & 3rd\\ \hline
spin $\downarrow$ & $-4.76$ & $-5.95$ & $-6.89$ & $-7.70$\\
spin $\uparrow$ & $-5.09$ & $-6.33$ & $-7.24$ & $-8.03$\\
difference & $-0.33$ & $-0.38$ & $-0.35$ & $-0.33$\\
\end{tabular}
\end{ruledtabular}
\end{table}

\section{Impurity solvers and convergence} \label{app:ed}
\subsection{CT-QMC and analytic continuation}
We solve the DMFT impurity problem at $T=0.01$ using the continuous-time hybridization-expansion solver~\cite{Werner2006}. To analytically continue each self-energy component, we subtract the static Hartree shift and construct the auxiliary Green's function~\cite{Wang2009}
\begin{equation}
G_{\rm aux}(i\omega_n)= [i\omega_n-\Sigma(i\omega_n)+\Sigma(\infty)]^{-1}.
\label{eq:aux}
\end{equation}
Because $G_{\rm aux}$ possesses a normalized fermionic spectral representation, it is directly amenable to maximum-entropy analytic continuation~\cite{Jarrell1996}. The real-axis self-energy is then recovered via $\Sigma(\omega)=\omega+\Sigma(\infty)-G_{\rm aux}^{-1}(\omega)$. The MaxEnt error parameter is set to $0.002$. In the strong-coupling Mott regime at half filling, rapid imaginary-time decay reaches the QMC statistical noise floor, necessitating the use of exact diagonalization.

\subsection{Real-axis ED bath construction}
For each spin component, the positive- and negative-frequency branches of the hybridization function are mapped onto two tight-binding chains using a moment-conserving Lanczos recursion~\cite{Si1994}. With $N_r$ bath sites per chain, the impurity problem contains $N_s=1+2N_r$ total sites. The hybridization is updated self-consistently directly on the real-frequency axis.

Lorentzian tails in the broadened hybridization density $\rho(\omega)=-\operatorname{Im}\Delta(\omega)/\pi$ can corrupt the high-order frequency moments $\int\omega^m\rho(\omega)\,d\omega$, which determine the parameters of the outer chain sites, displacing bath levels outside the physical band. To eliminate these spurious tails, we employ the Richardson extrapolation~\cite{Richardson1911} $2\rho_\eta-\rho_{2\eta}$, where $\rho_\eta=-\operatorname{Im}\Delta(\omega+i\eta)/\pi$. Because $\rho_\eta$ represents the bare hybridization convolved with a normalized Lorentzian $L_\eta(d)=(\eta/\pi)/(d^2+\eta^2)$, this linear combination evaluates to a convolution with the strictly positive kernel
\begin{equation}
2L_\eta(d)-L_{2\eta}(d)= \frac{6\eta^3/\pi}{(d^2+\eta^2)(d^2+4\eta^2)},
\label{eq:kernel}
\end{equation}
whose asymptotic tails decay as $d^{-4}$ rather than $d^{-2}$.

For large insulating baths, we remove residual broadening within the insulating gap, filter out spurious bath poles carrying less than $10^{-6}$ of the total hybridization weight, and truncate hybridization densities falling below $10^{-4}$ of the maximum. These precautions prevent the Lanczos recursion from generating unphysical in-gap states or spectral artifacts.

\subsection{Configuration-interaction truncation}
Full exact diagonalization is applied for baths up to $N_s=19$. Larger systems are solved within a truncated determinant space using the configuration-interaction approach~\cite{Zgid2011,Zgid2012,Lin2013,Lu2014,Go2017}. In the bath eigenbasis, the reference state fills all bath orbitals below the chemical potential and places a majority-spin electron on the impurity site. The variational space retains determinants hosting at most $2k$ particle--hole excitations relative to this reference for ground-state calculations, and $2k+1$ excitations for Green's functions, ensuring closure under electron addition and removal.

At $U=15$, setting $k=3$ reproduces full ED at $N_s=19$, matching edge-band weights to within $0.001$ and all twelve sideband peak energies within $0.01$. At $N_s=49$, the $k=2$ and $k=3$ spectra match to within $3\times10^{-7}$. We therefore adopt $k=2$ for large baths spanning $N_s=61$--$161$. At $N_s=161$, the $k=2$ subspace comprises $6.2\times10^7$ ground-state determinants and up to $2.8\times10^9$ Green's-function configurations.

At weaker interactions, larger excitation subspaces are required. We employ $k=4$ for $U=3$ ($N_s=33$), and $k=3$ for $U=4.5$, $6$, and $9$ ($N_s=61$). At $U=4.5$, $k=3$ reproduces $k=4$ at $N_s=33$ to within $10^{-4}$. Expanding the truncation at accessible bath sizes leaves the staggered moment and edge band unchanged. The deeper sidebands at $U=6$ require $k=4$, matching $k=5$ within $4\%$. Their detailed structure at $N_s=61$, $k=3$ is therefore not converged, which does not affect the conclusions about the edge splitting. The largest baths at $U=6$ and $9$ ($N_s=81$--$161$, $k=2$), used to extract the flat fraction (Fig.~\ref{fig:flat}), yield moments, gaps, and edge splittings in full agreement with $N_s=61$, $k=3$.

Table~\ref{tab:ed} summarizes the numerical convergence of edge observables at $U=15$. Across $N_s=49$--$161$, the edge quantities are converged, with $\DEc=0.32$--$0.34$ and $\Zedge=0.119$--$0.122$. The sideband positions remain unchanged from $N_s=81$ to $161$ (first and second sidebands $1.2$ and $2.1$ below the edge peak), whereas the finer oscillations proliferate with the number of bath levels and reflect the discretization.

\begin{table}[b]
\caption{Half-filled ED-DMFT convergence at $U=15$. $\Zedge$ is the spin-averaged edge-band weight per unit removal weight, integrated from $0.1$ below the peak to the gap. The refined bath construction is used for $N_s\ge49$. The larger-bath spectra are compared in Fig.~\ref{fig:afm}.}
\label{tab:ed}
\begin{ruledtabular}
\begin{tabular}{lccccc}
$N_s$ & $m_s$ & double occ. & $\Zedge$ & $\DEc$ & $\DEc/\Zedge$\\ \hline
17 (full ED) & 0.9729 & 0.0129 & 0.147 & 0.41 & 2.78\\
49 ($k=3$) & 0.9729 & 0.0129 & 0.119 & 0.32 & 2.67\\
61 ($k=2$) & 0.9729 & 0.0129 & 0.122 & 0.34 & 2.80\\
121 ($k=2$) & 0.9729 & 0.0129 & 0.120 & 0.34 & 2.85\\
141 ($k=2$) & 0.9729 & 0.0129 & 0.120 & 0.33 & 2.79\\
161 ($k=2$) & 0.9729 & 0.0129 & 0.120 & 0.33 & 2.79\\
\end{tabular}
\end{ruledtabular}
\end{table}

\bibliography{draft}

@article{Smejkal2022a,
  author  = {\v{S}mejkal, L. and Sinova, J. and Jungwirth, T.},
  title   = {Beyond conventional ferromagnetism and antiferromagnetism: A phase with nonrelativistic spin and crystal rotation symmetry},
  journal = {Phys. Rev. X},
  volume  = {12},
  pages   = {031042},
  year    = {2022},
  doi     = {10.1103/physrevx.12.031042}
}

@article{Smejkal2022b,
  author  = {\v{S}mejkal, L. and Sinova, J. and Jungwirth, T.},
  title   = {Emerging research landscape of altermagnetism},
  journal = {Phys. Rev. X},
  volume  = {12},
  pages   = {040501},
  year    = {2022},
  doi     = {10.1103/physrevx.12.040501}
}

@article{Krempasky2024,
  author  = {Krempask\'y, J. and others},
  title   = {Altermagnetic lifting of {Kramers} spin degeneracy},
  journal = {Nature},
  volume  = {626},
  pages   = {517},
  year    = {2024},
  doi     = {10.1038/s41586-023-06907-7}
}

@article{Reimers2024,
  author  = {Reimers, S. and others},
  title   = {Direct observation of altermagnetic band splitting in {CrSb} thin films},
  journal = {Nat. Commun.},
  volume  = {15},
  pages   = {2116},
  year    = {2024},
  doi     = {10.1038/s41467-024-46476-5}
}

@article{Guo2023,
  author  = {Guo, Y. and Liu, H. and Janson, O. and Fulga, I. C. and van den Brink, J. and Facio, J. I.},
  title   = {Spin-split collinear antiferromagnets: A large-scale ab-initio study},
  journal = {Mater. Today Phys.},
  volume  = {32},
  pages   = {100991},
  year    = {2023},
  doi     = {10.1016/j.mtphys.2023.100991}
}

@article{Wei2025,
  author  = {Wei, C.-C. and others},
  title   = {{La$_2$O$_3$Mn$_2$Se$_2$}: A correlated insulating layered {$d$}-wave altermagnet},
  journal = {Phys. Rev. Mater.},
  volume  = {9},
  pages   = {024402},
  year    = {2025},
  doi     = {10.1103/physrevmaterials.9.024402}
}

@article{Das2024,
  author  = {Das, P. and Leeb, V. and Knolle, J. and Knap, M.},
  title   = {Realizing altermagnetism in {Fermi-Hubbard} models with ultracold atoms},
  journal = {Phys. Rev. Lett.},
  volume  = {132},
  pages   = {263402},
  year    = {2024},
  doi     = {10.1103/physrevlett.132.263402}
}

@article{Giuli2025,
  author  = {Giuli, S. and Mejuto-Zaera, C. and Capone, M.},
  title   = {Altermagnetism from interaction-driven itinerant magnetism},
  journal = {Phys. Rev. B},
  volume  = {111},
  pages   = {L020401},
  year    = {2025},
  doi     = {10.1103/physrevb.111.l020401}
}

@article{Brinkman1970,
  author  = {Brinkman, W. F. and Rice, T. M.},
  title   = {Single-particle excitations in magnetic insulators},
  journal = {Phys. Rev. B},
  volume  = {2},
  pages   = {1324},
  year    = {1970},
  doi     = {10.1103/physrevb.2.1324}
}

@article{Kane1989,
  author  = {Kane, C. L. and Lee, P. A. and Read, N.},
  title   = {Motion of a single hole in a quantum antiferromagnet},
  journal = {Phys. Rev. B},
  volume  = {39},
  pages   = {6880},
  year    = {1989},
  doi     = {10.1103/physrevb.39.6880}
}

@article{Martinez1991,
  author  = {Martinez, G. and Horsch, P.},
  title   = {Spin polarons in the {$t$-$J$} model},
  journal = {Phys. Rev. B},
  volume  = {44},
  pages   = {317},
  year    = {1991},
  doi     = {10.1103/physrevb.44.317}
}

@article{Georges1996,
  author  = {Georges, A. and Kotliar, G. and Krauth, W. and Rozenberg, M. J.},
  title   = {Dynamical mean-field theory of strongly correlated fermion systems and the limit of infinite dimensions},
  journal = {Rev. Mod. Phys.},
  volume  = {68},
  pages   = {13},
  year    = {1996},
  doi     = {10.1103/revmodphys.68.13}
}

@article{Jarrell1996,
  author  = {Jarrell, M. and Gubernatis, J. E.},
  title   = {Bayesian inference and the analytic continuation of imaginary-time quantum {Monte Carlo} data},
  journal = {Phys. Rep.},
  volume  = {269},
  pages   = {133},
  year    = {1996},
  doi     = {10.1016/0370-1573(95)00074-7}
}

@article{Wang2009,
  author  = {Wang, X. and Gull, E. and de' Medici, L. and Capone, M. and Millis, A. J.},
  title   = {Antiferromagnetism and the gap of a {Mott} insulator: Results from analytic continuation of the self-energy},
  journal = {Phys. Rev. B},
  volume  = {80},
  pages   = {045101},
  year    = {2009},
  doi     = {10.1103/physrevb.80.045101}
}

@article{Werner2006,
  author  = {Werner, P. and Comanac, A. and de' Medici, L. and Troyer, M. and Millis, A. J.},
  title   = {Continuous-time solver for quantum impurity models},
  journal = {Phys. Rev. Lett.},
  volume  = {97},
  pages   = {076405},
  year    = {2006},
  doi     = {10.1103/physrevlett.97.076405}
}

@article{Caffarel1994,
  author  = {Caffarel, M. and Krauth, W.},
  title   = {Exact diagonalization approach to correlated fermions in infinite dimensions: {Mott} transition and superconductivity},
  journal = {Phys. Rev. Lett.},
  volume  = {72},
  pages   = {1545},
  year    = {1994},
  doi     = {10.1103/physrevlett.72.1545}
}

@article{Si1994,
  author  = {Si, Q. and Rozenberg, M. J. and Kotliar, G. and Ruckenstein, A. E.},
  title   = {Correlation induced insulator to metal transitions},
  journal = {Phys. Rev. Lett.},
  volume  = {72},
  pages   = {2761},
  year    = {1994},
  doi     = {10.1103/physrevlett.72.2761}
}

@article{Zgid2011,
  author  = {Zgid, D. and Chan, G. K.-L.},
  title   = {Dynamical mean-field theory from a quantum chemical perspective},
  journal = {J. Chem. Phys.},
  volume  = {134},
  pages   = {094115},
  year    = {2011},
  doi     = {10.1063/1.3556707}
}

@article{Lin2013,
  author  = {Lin, C. and Demkov, A. A.},
  title   = {Efficient variational approach to the impurity problem and its application to the dynamical mean-field theory},
  journal = {Phys. Rev. B},
  volume  = {88},
  pages   = {035123},
  year    = {2013},
  doi     = {10.1103/PhysRevB.88.035123}
}

@article{Zgid2012,
  author  = {Zgid, D. and Gull, E. and Chan, G. K.-L.},
  title   = {Truncated configuration interaction expansions as solvers for correlated quantum impurity models and dynamical mean-field theory},
  journal = {Phys. Rev. B},
  volume  = {86},
  pages   = {165128},
  year    = {2012},
  doi     = {10.1103/physrevb.86.165128}
}

@article{Lu2014,
  author  = {Lu, Y. and H\"oppner, M. and Gunnarsson, O. and Haverkort, M. W.},
  title   = {Efficient real-frequency solver for dynamical mean-field theory},
  journal = {Phys. Rev. B},
  volume  = {90},
  pages   = {085102},
  year    = {2014},
  doi     = {10.1103/physrevb.90.085102}
}

@article{Go2017,
  author  = {Go, A. and Millis, A. J.},
  title   = {Adaptively truncated {Hilbert} space based impurity solver for dynamical mean-field theory},
  journal = {Phys. Rev. B},
  volume  = {96},
  pages   = {085139},
  year    = {2017},
  doi     = {10.1103/physrevb.96.085139}
}

@article{Strack1992,
  author  = {Strack, R. and Vollhardt, D.},
  title   = {Dynamics of a hole in the {$t$-$J$} model with local disorder: Exact results for high dimensions},
  journal = {Phys. Rev. B},
  volume  = {46},
  pages   = {13852},
  year    = {1992},
  doi     = {10.1103/physrevb.46.13852}
}

@article{Sangiovanni2006,
  author  = {Sangiovanni, G. and Toschi, A. and Held, K. and Capone, M. and Gunnarsson, O. and Mo, S.-K. and Allen, J. W.},
  title   = {Static versus dynamical mean-field theory of {Mott} antiferromagnets},
  journal = {Phys. Rev. B},
  volume  = {73},
  pages   = {205121},
  year    = {2006},
  doi     = {10.1103/physrevb.73.205121}
}

@article{Daghofer2026,
  author  = {Daghofer, M. and Wohlfeld, K. and van den Brink, J.},
  title   = {Altermagnetic polarons: The fate of altermagnetic band splittings at strong coupling},
  journal = {Phys. Rev. Lett.},
  volume  = {136},
  pages   = {146502},
  year    = {2026},
  eprint  = {2506.03261},
  archivePrefix = {arXiv},
  doi     = {10.1103/261t-cs82}
}

@article{Lanzini2026,
  author  = {Lanzini, L. and Das, P. and Knap, M.},
  title   = {Theory of angle resolved photoemission spectroscopy of altermagnetic {Mott} insulators},
  journal = {Phys. Rev. Lett.},
  volume  = {137},
  pages   = {056501},
  year    = {2026},
  doi     = {10.1103/1d1n-cssv},
  eprint  = {2506.03263},
  archivePrefix = {arXiv}
}

@article{Leraand2026,
  author  = {Leraand, K. and M\ae{}land, K. and Sudb\o{}, A.},
  title   = {Spin-dependent quasiparticle lifetimes in altermagnets},
  journal = {Phys. Rev. B},
  volume  = {113},
  pages   = {115148},
  year    = {2026},
  doi     = {10.1103/r3vm-m1m3}
}

@article{Park2025,
  author  = {Park, I. and Birol, T. and Georges, A. and Fernandes, R. M.},
  title   = {Impact of strong electronic correlations on altermagnets: The case of {NiS$_2$}},
  journal = {Phys. Rev. Mater.},
  volume  = {10},
  pages   = {054415},
  year    = {2026},
  doi     = {10.1103/pgp6-zlh8}
}

@article{Kanamori1963,
  author  = {Kanamori, J.},
  title   = {Electron correlation and ferromagnetism of transition metals},
  journal = {Prog. Theor. Phys.},
  volume  = {30},
  pages   = {275},
  year    = {1963},
  doi     = {10.1143/ptp.30.275}
}

@article{Hirsch1985,
  author  = {Hirsch, J. E.},
  title   = {Two-dimensional {Hubbard} model: Numerical simulation study},
  journal = {Phys. Rev. B},
  volume  = {31},
  pages   = {4403},
  year    = {1985},
  doi     = {10.1103/physrevb.31.4403}
}

@article{Shastry1990,
  author  = {Shastry, B. S. and Krishnamurthy, H. R. and Anderson, P. W.},
  title   = {Instability of the {Nagaoka} ferromagnetic state of the {$U=\infty$} {Hubbard} model},
  journal = {Phys. Rev. B},
  volume  = {41},
  pages   = {2375},
  year    = {1990},
  doi     = {10.1103/physrevb.41.2375}
}

@article{Vollhardt1997,
  author  = {Vollhardt, D. and Bl\"umer, N. and Held, K. and Kollar, M. and Schlipf, J. and Ulmke, M.},
  title   = {Non-perturbative approaches to magnetism in strongly correlated electron systems},
  journal = {Z. Phys. B},
  volume  = {103},
  pages   = {283},
  year    = {1997},
  doi     = {10.1007/s002570050375}
}

@article{Richardson1911,
  author  = {Richardson, L. F.},
  title   = {The approximate arithmetical solution by finite differences of physical problems involving differential equations, with an application to the stresses in a masonry dam},
  journal = {Phil. Trans. R. Soc. Lond. A},
  volume  = {210},
  pages   = {307},
  year    = {1911},
  doi     = {10.1098/rsta.1911.0009}
}

@article{DelRe2025,
  author  = {{L. Del Re}},
  title   = {{Dirac} points and topological phases in correlated altermagnets},
  journal = {Phys. Rev. Research},
  volume  = {7},
  pages   = {033234},
  year    = {2025},
  doi     = {10.1103/7nvm-s225}
}

@article{Roig2024,
  author  = {Roig, M. and Kreisel, A. and Yu, Y. and Andersen, B. M. and Agterberg, D. F.},
  title   = {Minimal models for altermagnetism},
  journal = {Phys. Rev. B},
  volume  = {110},
  pages   = {144412},
  year    = {2024},
  doi     = {10.1103/PhysRevB.110.144412}
}

@article{Gondolf2025,
  author  = {Gondolf, J. and Kreisel, A. and Roig, M. and Yu, Y. and Agterberg, D. F. and Andersen, B. M.},
  title   = {Local signatures of altermagnetism},
  journal = {Phys. Rev. B},
  volume  = {111},
  pages   = {174436},
  year    = {2025},
  doi     = {10.1103/PhysRevB.111.174436}
}

@article{Metzner1989,
  author  = {Metzner, W. and Vollhardt, D.},
  title   = {Correlated lattice fermions in {$d=\infty$} dimensions},
  journal = {Phys. Rev. Lett.},
  volume  = {62},
  pages   = {324},
  year    = {1989},
  doi     = {10.1103/PhysRevLett.62.324}
}

@article{MullerHartmann1989,
  author  = {M\"uller-Hartmann, E.},
  title   = {Correlated fermions on a lattice in high dimensions},
  journal = {Z. Phys. B},
  volume  = {74},
  pages   = {507},
  year    = {1989},
  doi     = {10.1007/BF01311397}
}

@article{ARPES2,
  author  = {Yang, G. and others},
  title   = {Three-dimensional mapping of the altermagnetic spin splitting in {CrSb}},
  journal = {Nat. Commun.},
  volume  = {16},
  pages   = {1442},
  year    = {2025},
  doi     = {10.1038/s41467-025-56647-7}
}

@article{ARPES5,
  author  = {Lee, S. and others},
  title   = {Broken {Kramers} degeneracy in altermagnetic {MnTe}},
  journal = {Phys. Rev. Lett.},
  volume  = {132},
  pages   = {036702},
  year    = {2024},
  doi     = {10.1103/PhysRevLett.132.036702}
}

@article{jungwirth2025altermagnetism,
  author  = {Jungwirth, T. and Fernandes, R. M. and Fradkin, E. and MacDonald, A. H. and Sinova, J. and \v{S}mejkal, L.},
  title   = {Altermagnetism: {An} unconventional spin-ordered phase of matter},
  journal = {Newton},
  volume  = {1},
  pages   = {100162},
  year    = {2025},
  doi     = {10.1016/j.newton.2025.100162}
}

@article{jungwirth2026symmetry,
  author  = {Jungwirth, T. and others},
  title   = {Symmetry, microscopy and spectroscopy signatures of altermagnetism},
  journal = {Nature},
  volume  = {649},
  pages   = {837},
  year    = {2026},
  doi     = {10.1038/s41586-025-09883-2}
}

@article{hayami2019momentum,
  author  = {Hayami, S. and Yanagi, Y. and Kusunose, H.},
  title   = {Momentum-dependent spin splitting by collinear antiferromagnetic ordering},
  journal = {J. Phys. Soc. Jpn.},
  volume  = {88},
  pages   = {123702},
  year    = {2019},
  doi     = {10.7566/JPSJ.88.123702}
}

@article{Maier2005,
  author  = {Maier, Thomas and Jarrell, Mark and Pruschke, Thomas and Hettler, Matthias H.},
  title   = {Quantum cluster theories},
  journal = {Rev. Mod. Phys.},
  volume  = {77},
  pages   = {1027},
  year    = {2005},
  doi     = {10.1103/RevModPhys.77.1027}
}

@article{Rohringer2018,
  author  = {Rohringer, G. and Hafermann, H. and Toschi, A. and Katanin, A. A. and Antipov, A. E. and Katsnelson, M. I. and Lichtenstein, A. I. and Rubtsov, A. N. and Held, K.},
  title   = {Diagrammatic routes to nonlocal correlations beyond dynamical mean field theory},
  journal = {Rev. Mod. Phys.},
  volume  = {90},
  pages   = {025003},
  year    = {2018},
  doi     = {10.1103/RevModPhys.90.025003}
}

@article{Trugman1988,
  author  = {Trugman, S. A.},
  title   = {Interaction of holes in a {H}ubbard antiferromagnet and high-temperature superconductivity},
  journal = {Phys. Rev. B},
  volume  = {37},
  pages   = {1597},
  year    = {1988},
  doi     = {10.1103/PhysRevB.37.1597}
}

@article{Werner2007,
  author  = {Werner, Philipp and Millis, Andrew J.},
  title   = {Doping-driven {M}ott transition in the one-band {H}ubbard model},
  journal = {Phys. Rev. B},
  volume  = {75},
  pages   = {085108},
  year    = {2007},
  doi     = {10.1103/PhysRevB.75.085108}
}

\end{document}